\documentclass[final,3p, 12pt, authoryear]{elsarticle}
\journal{XX}
\makeatletter
\def\ps@pprintTitle{%
	\let\@oddhead\@empty
	\let\@evenhead\@empty
	\def\@oddfoot{}%
	\let\@evenfoot\@oddfoot}
\makeatother

\usepackage[german,english]{babel}
\usepackage[utf8]{inputenc}
\usepackage[T1]{fontenc}
\usepackage[german=guillemets,autostyle = true]{csquotes}
\MakeAutoQuote{«}{»}
\usepackage{graphicx}
\usepackage{amsmath}
\usepackage{amssymb}
\usepackage{amsfonts}
\usepackage{amsthm}
\usepackage{bm}
\usepackage{arydshln}
\usepackage{bbm} 
\usepackage{mathrsfs}
\usepackage[mathcal]{euscript}
\usepackage{mathtools}
\usepackage{mathptmx}

\usepackage{xcolor}
\usepackage[deletedmarkup=xout,highlightmarkup=background,authormarkup=none,final]{changes}

\usepackage{verbatim}
\usepackage{enumitem}

\usepackage{arydshln}

\usepackage{lipsum} 

\usepackage{stackengine}
\usepackage{placeins}
\usepackage{eurosym}

\usepackage{textgreek}

\setquotestyle{english}

\addto\captionsenglish{}
\usepackage[position=b]{subcaption}

\usepackage{lineno}

\usepackage{titlesec}
\usepackage{hyperref}
\usepackage{cleveref}

\titleclass{\subsubsubsection}{straight}[\subsection]
\newcounter{subsubsubsection}[subsubsection]
\renewcommand\thesubsubsubsection{\thesubsubsection.\arabic{subsubsubsection}}
\titleformat{\subsubsubsection}{\normalfont\normalsize\itshape}{\thesubsubsubsection}{1em}{}
\titlespacing*{\subsubsubsection}{0pt}{3.25ex plus 1ex minus .2ex}{1.5ex plus .2ex}
\makeatletter
\def\toclevel@subsubsubsection{4}
\def\l@subsubsubsection{\@dottedtocline{4}{7em}{4em}}
\makeatother
\usepackage{tikz}
\usetikzlibrary{shapes.arrows}

\newcommand\myprime{\mkern-3.5mu\raise0.6ex\hbox{$\scriptstyle\prime$}}

\begin{document}
	\graphicspath{{./Figures/}}
	
	\def\myDownArrow{\smash{
			\begin{tikzpicture}[baseline=-2mm]
				\useasboundingbox (-2,0);
				\node[single arrow,draw=black,fill=black!10,minimum height=1cm,shape border rotate=270] at (0,-1) {};
			\end{tikzpicture}
	}}
	\def\myRightArrow{\smash{
			\begin{tikzpicture}[baseline=-1mm]
				\useasboundingbox (-2,0);
				\node[single arrow,draw=black,fill=black!10,minimum height=2cm,shape border rotate=0] at (0,-1) {};
			\end{tikzpicture}
	}}
	\begin{frontmatter}
		
		\title{A unifying review of NDE models towards optimal decision support}
		
		\author[ERA]{Elizabeth Bismut\corref{cor1}}\ead{elizabeth.bismut@tum.de}
		\author[ERA]{Daniel Straub}\ead{straub@tum.de}
		
		\cortext[cor1]{Corresponding author.} 
	
	\address[ERA]{Engineering Risk Analysis Group, Technische Universität München.\\ Arcisstra{\ss}e 21, 80290 München, Germany}
	
	\begin{abstract}
		Non-destructive evaluation (NDE)  inspections  are an integral part of asset integrity management. The relationship between the condition of interest and the quantity measured by NDE is described with probabilistic models such as PoD or ROC curves. These models are used to assess the quality of the information provided by NDE systems, which is affected by factors such as the experience of the inspector, environmental conditions, ease of access, and the precision of the measurement device. In this paper, we review existing probabilistic models of NDE and show how they are connected within a unifying framework.  This framework provides insights into how these models should be learned, calibrated, and applied. We investigate and highlight how the choice of the model can affect the maintenance decisions taken on the basis of NDE results. In addition, we analyze the impact of experimental design on the performance of a given NDE system in a decision-making context.

	\end{abstract}
	
	
\end{frontmatter}
\newpage

\section{Introduction}\label{Sec:Intro}
Probabilistic models  have been developed to measure the quality and performance of non-destructive evaluation (NDE) methods. 
These models include probability of detection (PoD) curves, receiver operating characteristic (ROC) curves, or simply a probability of false positives/false negatives. 
In specific application domains, one of these models typically prevails \citep[e.g.,][]{Packman_Pearson_68,Hovey_Berens_88,Somoza_et_al_90,Sarkar_et_al_98}, since they historically emerged independently in different disciplines to address specific problems. 
PoD curve models were formulated in the 1950s as a statistical method to investigate the dose-response effect in biological tests \citep{Finney_78,Rudemo_et_al_87,Ritz_et_al_15}. 
ROC curve models grew out of the signal detectability theory developed in the 1940s \citep{Shannon_48, Woodward_Davies_52, Peterson_Birdsall_53}, aimed at measuring the capacity of a receiver to distinguish the presence of a signal from noise. 
The purpose of this paper is to review and investigate these models for NDE quality by means of a unifying framework, which shows formally  how they are connected. This enables us to provide insights into how they should be used and learned.

Historically, a first major application of NDE was the identification of flaws with systematic inspections during the manufacturing process of parts. This was especially a concern in the nuclear industry, which strived to improve quality assurance in the fabrication of pressure vessel components \citep{NDT_Nuclear_Symposium_65}. In the aeronautic industry, the high cost associated with discarding parts with small defects during manufacture motivated the use of NDE also during the service life. By the 1970s, the US Air Force had launched detailed investigations on quantifying the performance (also called reliability) of defect detection measures \citep{Packman_Pearson_68,NMAB-252_69}. Its program "Have-Cracks-Will-Travel" laid the foundation for systematic inspection and maintenance  of aircraft \citep{Berens_Hovey_81,Hovey_Berens_88,Singh_99}, where fatigue cracks are allowed to develop as long as they are monitored and repaired regularly. These studies provide the analytical framework to derive PoD curves from NDE data. 

In parallel, the research on lifetime extension of aircraft structures \citep{Graham_Tetelman_74,Yang_Trapp_74}, nuclear reactor components \citep{Harris_Lim_83}, and large deteriorating civil infrastructures \citep{Frangopol_et_al_97,Hong_97,Sheils_et_al_12} started to incorporate these probabilistic NDE models into Bayesian reliability analysis of the structure. This was also coupled with inspection and maintenance plans, notably in risk-based inspection (RBI) planning \citep{Yang_Chen_85,Madsen_et_al_87, Straub_04}. In the context of highly reliable structures, optimizing the time and location of an inspection and the eventual subsequent repair can result in significant savings \citep{Nielsen_Sorensen_10,Goulet_et_al_15,Luque_Straub_17,Bismut_Straub_19}. This explains why NDE studies now also focus on quantifying the added value of a specific NDE method on the expected total life-cycle cost.

The accuracy and reliability of NDE depends on environmental conditions (humidity, temperature, experimental setting) and inspector expertise, among other factors \citep[e.g.,][]{Lentz_et_al_02}. NDE quality models establish a probabilistic relationship between the condition of the structure and the testing outcome (or outcome assessment). They can be used to rate NDE methods and ensure their compliance with norms and standards in place \citep[e.g.,][]{DIN_25435_6,R_08_129_SKB}. 

In this contribution, we give an overview of the existing types of models for NDE quality, and show how they are connected within a unifying framework. While some existing literature does discuss relationships between two NDE quality models, we are not aware of any previous attempt to formalize the relationship between all these models. 
This framework enables a better understanding of the assumptions associated with each model type and, as we demonstrate, can help in obtaining better models for NDE quality for specific applications. We particularly draw attention to the dependence of the NDE quality model on the experimental design and the risks of misinterpreting and generalizing quality indicators of NDE quality models from experimental studies. 
This formalization allows us to identify systematically issues and potential pitfalls when establishing and applying these models. Decision analysis and the concept of value of information (VoI) \citep{Raiffa_Schlaifer_61} shed further light on what the implications are of not choosing the optimal model. The joint optimization of the calibration of the NDE system and the repair actions is formally addressed. 

After some definitions in \Cref{SubS:Mon_Model}, \Cref{SubS:Mon_type} introduces the framework and the models and their connections are reviewed and discussed in \Cref{Subs:SX,Subs:IY,Subs:IX,Subs:SY}. Considerations on model availability and uncertainty are made in \Cref{Subs:Model_Uncertainty}. \Cref{Sec:DecisionAnalysis} focuses on the exploitation of the NDE data within decision analysis. The findings are illustrated with two examples: In \Cref{SubS:Example_3and4}, we apply the framework and solve a basic decision problem for a purely theoretical NDE system; In \Cref{SubS:Example_1and2} we consider a real NDE technique, the half-cell potential, and we analyze the effect of model choice on the outcomes of optimal decision making for a one-step and a two-step decision problem. 

\section{A unifying framework for NDE quality} \label{Sec:Theory}
\subsection{NDE systems} \label{SubS:Mon_Model}

\begin{figure}[htbp]
	\centering
	\includegraphics[width=0.6\linewidth]{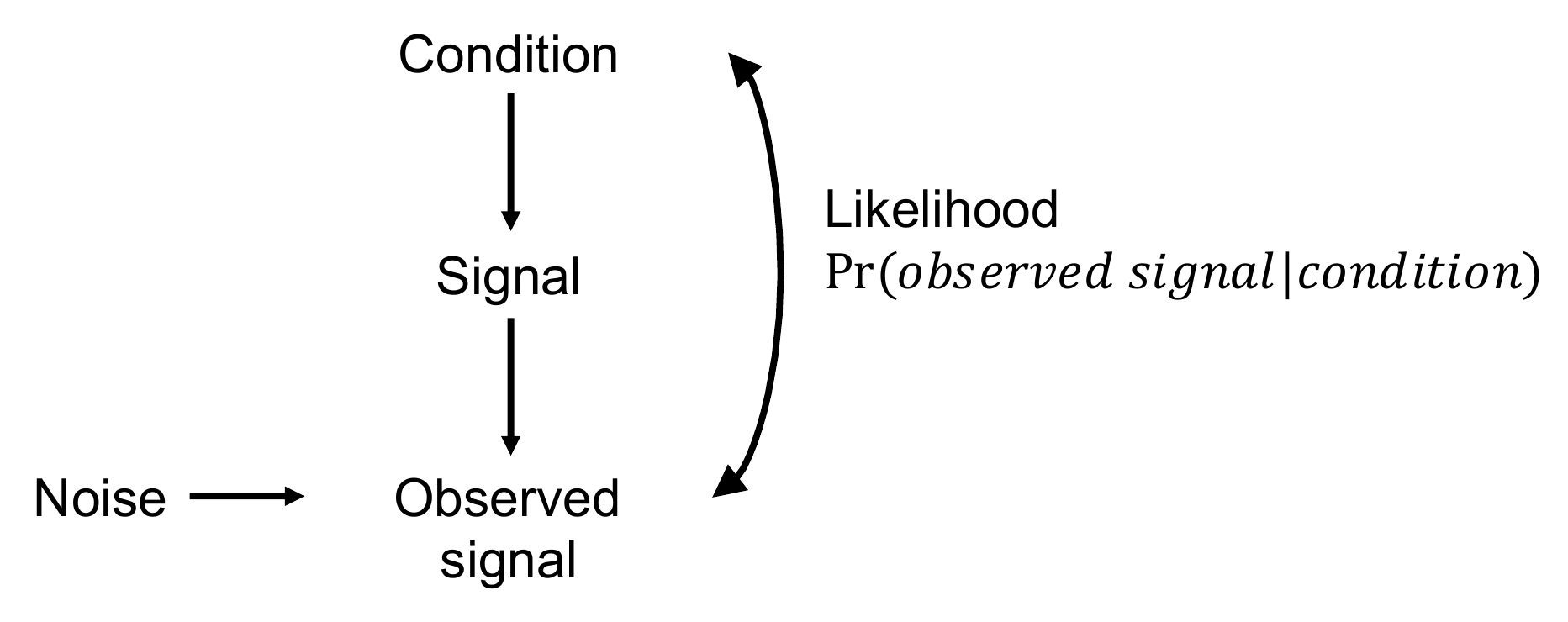}
	\caption{The observed signal measures the signal emitted by the condition and is also affected by noise. The model for NDE quality expresses the relationship between the observed signal and the condition, in the form of a likelihood.}
	\label{Fig:NDE_likelihood}
\end{figure}

NDE quality models establish a relationship between the true \emph{condition} and the measurement, which we call the \emph{observed signal}. The latter includes noise. Fig.~\ref{Fig:NDE_likelihood} depicts this relationship, inspired by Shannon's diagram of a communication system \citep{Shannon_48}.

The relationship between the true condition and the observed signal can be expressed probabilistically with \emph{NDE quality models}. Here the term "NDE" covers all non-destructive information collection methods, ranging from visual inspection to automated data collection.
An NDE quality model is defined by the conditional probability $\Pr(observed~signal|condition)$, which in statistics is known as the likelihood function.
This relationship can be derived empirically by performing a number of tests \citep{Packman_Pearson_68,Berens_Hovey_81,Berens_Vol1_00}.
While the primary purpose of these models was to establish a measure of reliability and performance of NDE techniques, they can be used for Bayesian analysis and  decision analysis, see Section~\ref{Sec:DecisionAnalysis}.

The \emph{NDE system} encapsulates the process of collecting the data (type of data collected, inspection technique) and the interpretation of this data \citep{Berens_Vol1_00}. 
The NDE system can encompass several measuring devices and data processors or interpreters \citep{Sheils_et_al_10}. The observed signal is the quantity on which the repair decision is taken. An example of an NDE system is an inspector going on-site, visually inspecting a wall, and appraising its state of damage (here, the condition) with a rating (here, the observed signal) \citep{Quirk_et_al_18}. The NDE quality model characterizes the NDE system.   

\subsection{Unifying NDE quality models}\label{SubS:Mon_type}
The NDE quality model probabilistically relates the observed signal with the condition. The condition can take continuous values (e.g., a crack size), or discrete values (e.g., "functioning" or "not functioning"). The observed signal can similarly take continuous (e.g., maximum vibration amplitude, measured crack length) or discrete values (e.g., "red" or "green", "suitable" or "not suitable").
Here, we limit the consideration of the discrete case to binary states. As the number of discrete states increases, the multinomial case approaches the continuous state. 

We denote the continuous observed signal by $S$ and the binary observed signal by $I$. 
Similarly, if the condition takes continuous values, it is denoted by $X$, and when it takes binary states, by $Y$. The condition can express a degree of damage or failure, although this remains an abstract concept and might not be related to anything failing as such. Here, the observed signal is considered as a scalar quantity, which can result from processing a multivariate signal, e.g., a time-series or an image \citep{Kurz_et_al_12,Webb_et_al_15}. 
Table~\ref{Tab:Monitor_models} gives an overview of the four main NDE quality model categories for the possible combinations of continuous or binary condition $X$/$Y$ and observed signal $S$/$I$. 
\renewcommand{\arraystretch}{1.8}
\begin{table}[!h]
	\caption{Monitoring models for binary or continuous signal and condition}
	\centering
	\renewcommand{\arraystretch}{0.9}
	\begin{tabular}{c c |c | c}
		& &\multicolumn{2}{c}{Condition}\\
		& &Continuous $X$ & Binary $Y$\\
		\hline
		Signal& 	\begin{tabular}{c}	
			Continuous $S$\\ \hline Binary $I$\end{tabular}  & \begin{tabular}{c}	(1) $f_{S|X=x}(s)$ \\ \hline (2)~~~$PoD(x)$ curve \end{tabular} &\begin{tabular}{c} (3) ROC curve\\ \hline (4) $PoD$ / $PFA$ \end{tabular}\\	
	\end{tabular}
	\label{Tab:Monitor_models}
\end{table}

The connection between the models comes from the fact that \emph{the binary/discrete variables are the result of imposing one or more thresholds on the underlying continuous variables}. Specifically, a binary signal $I$ is the result of a classification of the underlying continuous signal $S$, by means of a threshold $s_{th}$, that assigns $I=1$ for $S>s_{th}$.  A binary condition $Y$ represents two domains of a continuous condition $X$, classified such that $Y=1$ when $X>x_{th}$. 

\begin{figure*}[!h]
	
	\includegraphics[width=0.99\linewidth]{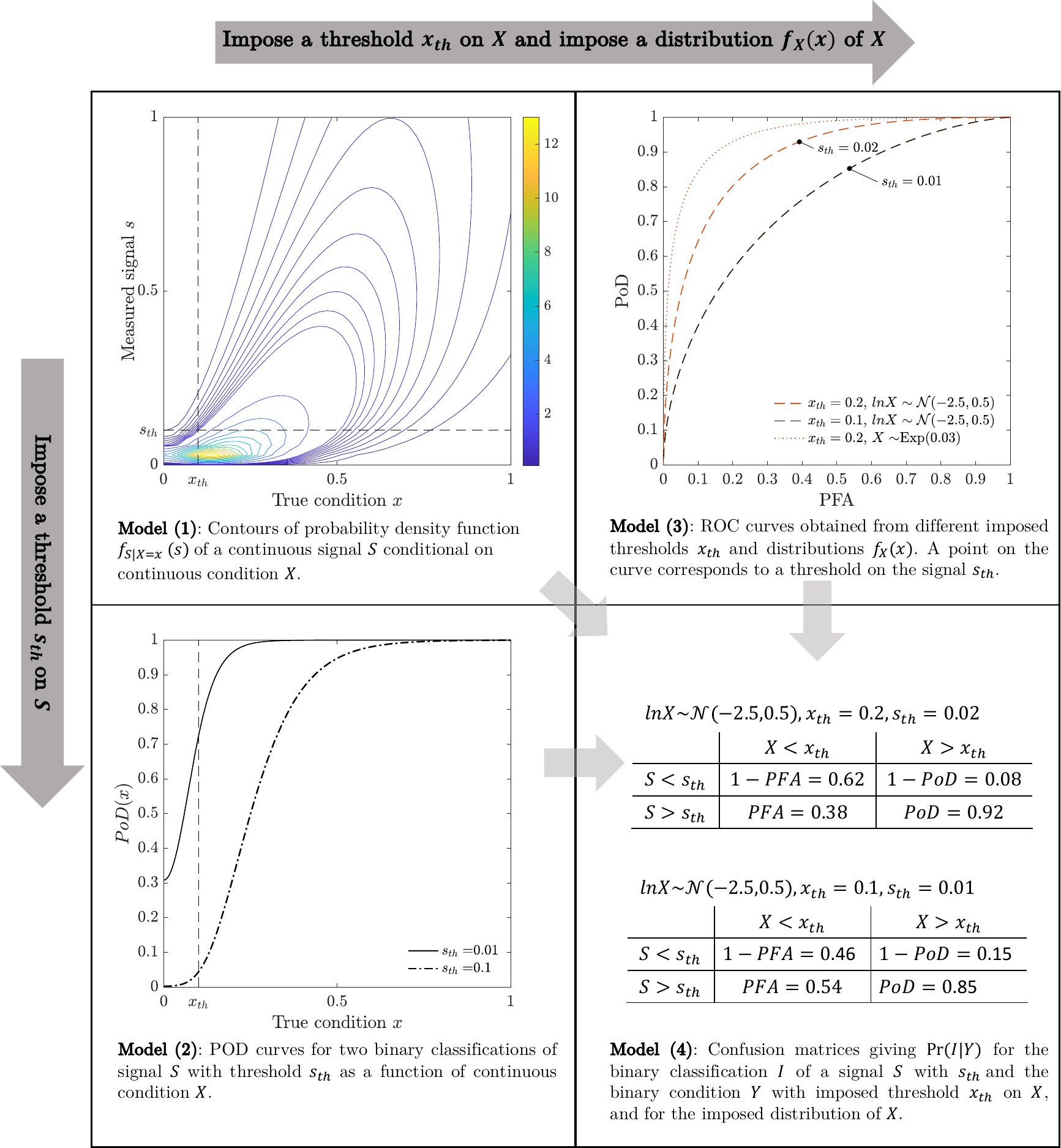}
		\caption{Unifying framework of models for NDE quality. Models (2--4) with a binary signal or condition can in principle be linked to Model (1). This link is established by fixing a threshold either on the signal, $s_{th}$, or on the condition, $x_{th}$, to classify the continuous signal or condition into binary states. The link between Models (1--2) to (3--4) requires additional information on the distribution of the continuous condition $X$. The specific NDE quality model depicted here is defined in \Cref{SubS:Base_hypothetical}.}
		\label{Fig:unified_model}
	\end{figure*}

If one has access to the full continuous/continuous description of NDE quality, one can establish the link between all four models, as illustrated by Fig.~\ref{Fig:unified_model}. Model (1) is the \emph{base model}. Models (2--4) are derived by imposing thresholds $s_{th}$ and $x_{th}$ on continuous signal $S$ and condition $X$, respectively. As we discuss in \Cref{Subs:Model_Uncertainty}, almost all NDE systems and applications can be connected to the base model.

	The mathematical formulation of this unifying framework is derived in Sections~\ref{Subs:SX} to~\ref{Subs:IY}. Importantly, we show that the links between models require an understanding of the population of defects in the experimental design. This affects the validity of the NDE models outside of the experimental setting in which they are learned. This has ramifications on the optimal interpretation of data and the decisions taken based on NDE.
	
	\subsection{Model (1): Base model -- $S$ continuous, $X$ continuous}\label{Subs:SX}
	Many NDE methods relate a continuous condition to a continuous observed signal. For example, ultrasonic testing (UT) detects discontinuities inside a metal plate by emitting a high frequency ultrasonic pulse towards the plate and recording the echo. The amplitude of this echo (the observed signal) relates to the thickness (the condition) of defect-free material \citep{Lavender_76}. 
	Another example described by this model is crack detection and measurement using magnetic particle inspection (MPI): MPI reveals the crack to the inspector who must then perform a visual inspection under UV lights and evaluate the crack size \citep{Clark_et_al_87}. 
	
	In the configuration of Model (1), the NDE system is fully characterized by the probability distribution of the observed signal $S$ given the condition $X$, through the conditional probability density function (pdf), $f_{S|X}(s|x)$, or the associated cumulative distribution function (cdf), $F_{S|X}(s|x)$. Fig.~\ref{Fig:unified_model} gives an example of such a conditional pdf.

	\replaced{This}
	{Typically, the} probabilistic model can be obtained from experimental data (experimental test blocks), if possible in different experimental settings. A traditional approach, which would allow to find such a model, is called "\textit{$\hat a$ vs. $a$,}" where $\hat a$ is the continuous observed signal and  $a$ the continuous condition \citep{Berens_Vol1_00}. A relationship of the form $\hat a=f(a)+\epsilon(a)$ has been proposed, where $f$ is the mean response function with some fixed parameters and $\epsilon(a)$ is a random variable representing the measurement noise. 
	A linear log-logistic relationship is a common choice for the function $f$ \citep{Berens_Hovey_81,MIL-HDBK-1823A_09}. It originates from biological tests investigating the dose-response effect \citep{Finney_78,Rudemo_et_al_87,Ritz_et_al_15}. $\epsilon(a)$ is commonly modeled as a Gaussian random variable \citep[e.g.,][]{Kurz_et_al_12,Goulet_et_al_15}. \citet{MIL-HDBK-1823A_09} provides some guidance as to how the noise should be considered. 

	Simulation and meta-models of NDE processes have also given rise to model-assisted PoD (MAPOD) \citep{Aldrin_et_al_13,Calmon_12,Calmon_et_al_16}. In this procedure, physics-based models are used to determine the relationship $f$ and are validated with the experimental data. 
	One of the advantage of MAPOD is that it does not require large datasets and can include numerous experimental settings and influential parameters.

	
	Our review of the existing literature on NDE models shows that continuous/continuous probabilistic models are often learned in an ad-hoc manner. An example is the probability of (correct) sizing (POS), which describes the error in the measurement by an inspector of the continuous condition (e.g., a crack length) 
	\citep{Brennan_13,daSilva_dePadua_12,Granville_Charlton_16,Visser_18,Nath_21}. Models for POS are continuous/continuous, but definitions vary and no application to reliability analysis is documented in the literature. 

For many NDE techniques, this model remains abstract, as a continuous signal or a continuous condition might not be easily identifiable. In this case, the NDE quality model is chosen among the other three categories described below. 

\subsection{Model (2): PoD curve -- $I$ binary, $X$ continuous}\label{Subs:IX}
The probability of detection curve, or \emph{$PoD$ curve}, has been adopted for many NDE techniques, such as UT, MPI, Eddy current testing, or impulse radar, which aim at identifying cracks or more generally defects in structures \citep{Berens_Hovey_81,Hovey_Berens_88, Sarkar_et_al_98,Feistkorn_Taffe_11}. It is
\begin{equation}\label{Eq:DefPoD}
	PoD(x)=\Pr(I=1|X=x).
\end{equation}

As previously noted, one can interpret $I$ as a classification of a continuous signal $S$ by fixing a threshold $s_{th}$, i.e., $\{I=1\}=\{S>s_{th}\}$. In NDE literature pertaining to PoD curves, $s_{th}$ is called the decision or detection threshold \citep{Berens_Hovey_81,Sheils_et_al_12}. The PoD function in Eq.~\eqref{Eq:DefPoD} can thus be written as a function of the continuous/continuous Model (1).

\begin{equation}\label{Eq:Model_1_to_2}
	PoD(x)=\Pr(S>s_{th}|X=x)=\int_{s_{th}}^{+\infty}f_{S|X}(s|x)\mathrm{d}s=1-F_{S|X}(s_{th}|x).
\end{equation}

By changing the threshold $s_{th}$, the PoD curve changes \citep{Sarkar_et_al_98}; at the limit, it is
$PoD(x,s_{th}=-\infty)=1$ and $PoD(x,s_{th}=+\infty)=0$ for any value $x$.

PoD curves have indeed been obtained from "\textit{$\hat a$ vs. $a$}" models \citep{Berens_Vol1_00,MIL-HDBK-1823A_09,Virkkunen_et_al_19}. In aircraft integrity management, the threshold $s_{th}$ is chosen so that a critical crack size fixed by expert judgment \citep{Wood_Engle_79} is detected with a 90\% probability, with a 95\% confidence level \citep{Berens_Vol1_00}. According to \cite{Wood_Engle_79}, the basis for these target probability and confidence level is arbitrary and relates to a required degree of conservatism 
and to the practical implementation of NDE testing programs. The characterization of a PoD curve with this 90/95 target value is considered best practice, until today \citep[e.g.,][]{MIL-HDBK-1823A_09,Tschoeke_et_al_21}. However, as we show below in Section~\ref{Sec:DecisionAnalysis}, calibrating the PoD curve and fixing $s_{th}$ in this way without a comprehensive decision analysis may not ensure that the test is exploited to its full potential. Furthermore, although the term decision threshold originally refers to the fact that a repair action systematically follows the detection of a defect with $S>s_{th}$, it may not be optimal to do so in all circumstances.





Eq.~\eqref{Eq:Model_1_to_2} does not preclude the PoD curve from taking a non-zero value when $x=0$. A PoD curve for which $PoD(0)=0.3$ is depicted in Fig.~\ref{Fig:unified_model} for $s_{th}=0.01$. \cite{Straub_04} notes that PoD curves typically proposed in the literature, such as those obtained from the log-logistic model mentioned in Section~\ref{Subs:SX}, pass through the point $PoD(0)=0$, and thus do not include the possibility of false detection. \citep{Packman_Pearson_68,Berens_Hovey_81} have motivated this choice with damage tolerant design philosophy by presuming that unnecessary repairs can only improve the reliability of the system. However, \cite{Heasler_Doctor_96} argues that for risk-based decision analysis it is more appropriate to consider PoD models for which $PoD(0)>0$ since unnecessary repairs lead to additional costs. \cite{Straub_04} has proposed the term probability of indication (PoI) for PoD curves that include false alarms, i.e., those for which $PoD(0)>0$, which include the effect of noise on the indication of the condition. Nevertheless, we use the term PoD curve throughout this paper.

\subsection{Model (3): ROC curve -- $S$ continuous, $Y$ binary }\label{Subs:SY}
This model is commonly used to describe how the continuous observed signal $S$ from NDE can be interpreted to discriminate between the absence ($Y=0$) and presence ($Y=1$) of a flaw. Most traditional NDE methods can be described by this model \citep{Olin_Meeker_96}, which can also be used to compare the ability of inspectors (human or machine) to interpret NDE results \citep{Swets_83}. Recently, image processing techniques for crack detection and monitoring systems of pipe corrosion have been evaluated with this model \citep{Pakrashi_et_al_10, Jarvis_et_al_18}.

\begin{figure}[htbp]
	\centering
	\includegraphics[width=0.55\linewidth]{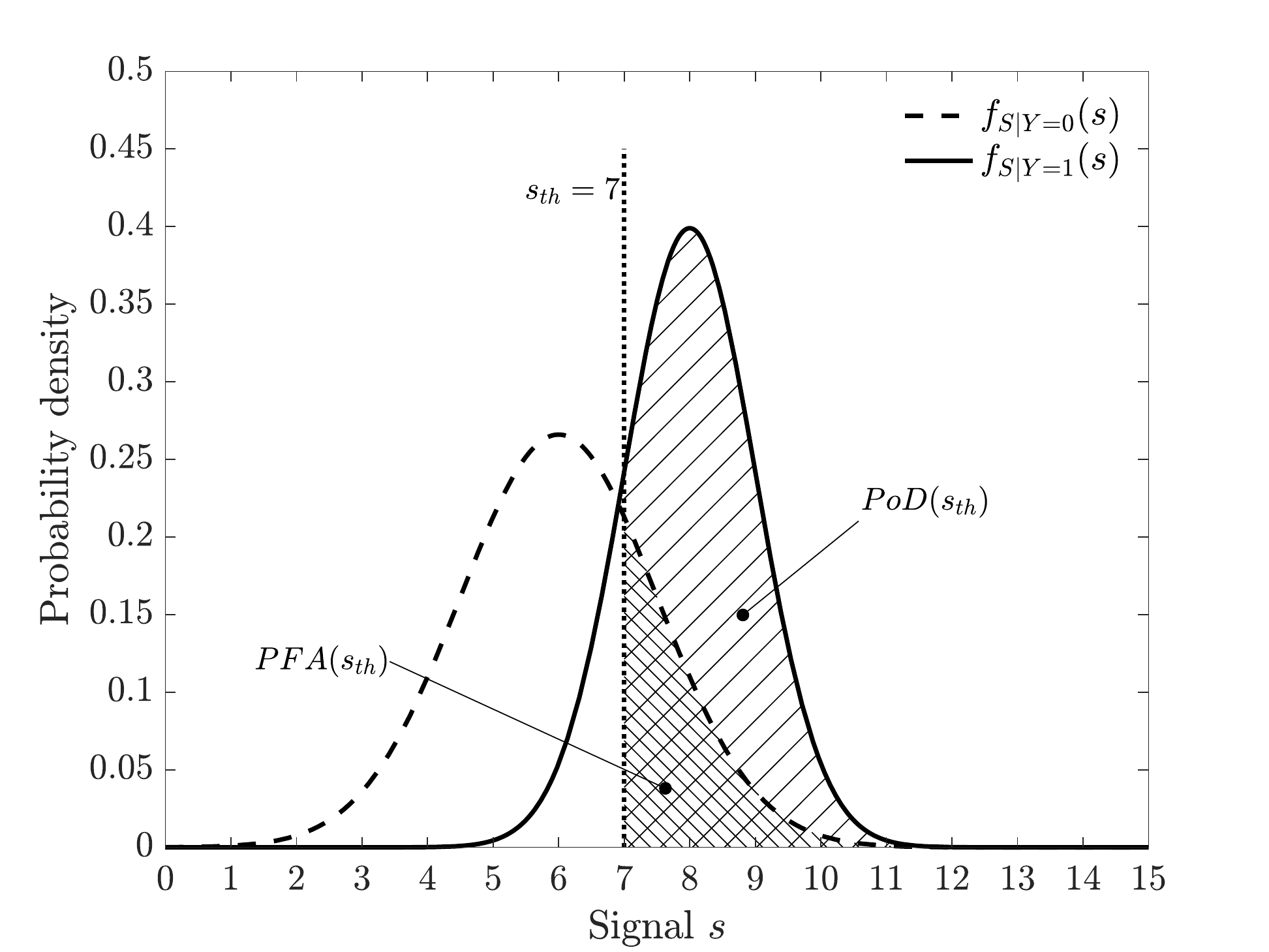}
	\caption{Conditional probability density functions. The indicated areas under the two curves defined by the threshold $s_{th}$ corresponds to a point $PoD(s_{th}=7)$, $PFA(s_{th}=7)$ on the ROC curve shown in Fig.~\ref{Fig:ROC_illustration}.}
	\label{Fig:pdfs_illustration}
\end{figure}

In this framework, the discrete condition $Y$ is defined by setting a threshold on the continuous condition $X$ such that $\{Y=1\}=\{X>x_{th}\}$ and $\{Y=0\}=\{X\leq x_{th}\}$. $x_{th}$ is called the critical threshold \citep{Sheils_et_al_10,Schoefs_et_al_12}. 
The likelihood function is formed by the two conditional pdfs of the observed signal $S$. They are represented in Fig.~\ref{Fig:pdfs_illustration} and can be derived from the base Model (1):

\begin{align}\label{Eq:Mod_1to3_fail}
	f_{S|Y=1}(s)=\frac{1}{1-F_{X}(x_{th})}\cdot\int_{x_{th}}^{+\infty}{f_{S|X}(s|x)f_{X}(x)\mathrm{d}x},\\\label{Eq:Mod_1to3_safe}
	f_{S|Y=0}(s)=\frac{1}{F_{X}(x_{th})}\cdot\int_{-\infty}^{x_{th}}{f_{S|X}(s|x)f_{X}(x)\mathrm{d}x},
\end{align}
where $f_{X}(x)$ and $F_{X}(x)$ are the pdf and cdf of the condition $X$.

This model is commonly visualized by the corresponding  receiver (or relative) operating characteristic (ROC) curve, which plots the $PoD$ against the $PFA$. This curve is parametrized by a threshold on the signal $s_{th}$, also called the cut-off point \citep{Fluss_et_al_05}. Note that the $PoD$ is not expressed as a function of the continuous condition $X$ as in Section~\ref{Subs:IX} above, but is here a function of $s_{th}$. 
The $PoD$ and $PFA$ as a function of $s_{th}$ are

\begin{align}\label{Eq:PoD_simple}
	PoD(s_{th})=\Pr(S>s_{th}|Y=1)=\int_{s_{th}}^{+\infty}{f_{S|Y=1}(s)\mathrm{d}s},\\\label{Eq:PFA_simple}
	PFA(s_{th})=\Pr(S>s_{th}|Y=0)=\int_{s_{th}}^{+\infty}{f_{S|Y=0}(s)\mathrm{d}s}.
\end{align}

\begin{figure}[!h]
	\centering
	\includegraphics[width=0.55\linewidth]{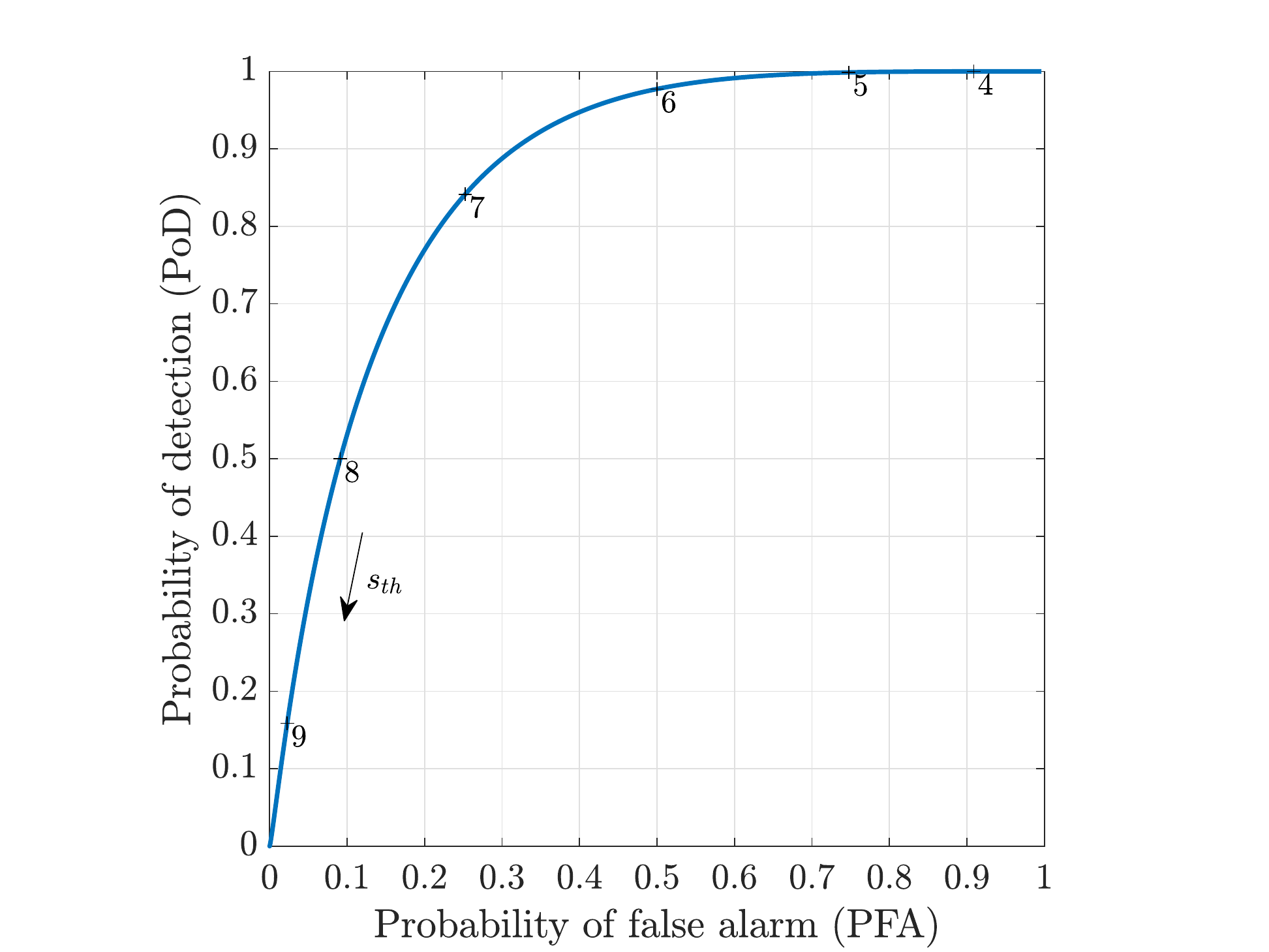}
	\caption{Each point on the ROC curve corresponds to a value of the signal threshold $s_{th}$.}
	\label{Fig:ROC_illustration}
\end{figure}

The ROC curve is illustrated in Fig.~\ref{Fig:ROC_illustration}, wherein the $PoD$ and $PFA$ for threshold values $s_{th}$ are obtained from the conditional pdfs of Fig.~\ref{Fig:pdfs_illustration}. Note that studies often derive the ROC curve for a single flaw size $X=x$, rather than for a domain defined by threshold $x_{th}$ \citep[e.g.,][]{Rajesh_93}. This can be useful to qualitatively compare NDE methods on a standardized flaw size but does not provide a complete model linking the observed signal with the condition.

One can alternatively express the $PoD$ and $PFA$ as a function of the conditional cdf $F_{S|X}$ from Model (1):

\begin{align}\label{Eq:PoD}
	PoD(s_{th})=\Pr(S>s_{th}|X>x_{th})=\frac{1}{1-F_{X}(x_{th})}\cdot\int_{x_{th}}^{+\infty}{\left(1-F_{S|X}(s_{th}|x)\right)f_{X}(x)\mathrm{d}x},
	\\\label{Eq:PFA}
	PFA(s_{th})=\Pr(S>s_{th}|X<x_{th})=\frac{1}{F_{X}(x_{th})}\cdot\int_{-\infty}^{x_{th}}{\left(1-F_{S|X}(s_{th}|x)\right)f_{X}(x)\mathrm{d}x}.
\end{align}

Several indicators have been proposed to qualify the performance of an NDE system with a ROC curve \citep{Taner_Antony_00,Greiner_et_al_00,Fluss_et_al_05}. They include the  \textit{area under the curve} (AUC), the shortest distance between the curve and $(0,1)$, also expressed in polar coordinates \citep{Schoefs_et_al_12}, and the \textit{Youden Index} \citep{Youden_50}. The latter is computed as the maximum vertical distance between the ROC curve and the 45 degree line starting at $(0,0)$.

\begin{figure}[!h]
\centering
\begin{subfigure}{0.5\linewidth}
	\centering
	\small Experimental designs\\
	\includegraphics[scale=0.45, trim=0 0 130 20,clip]{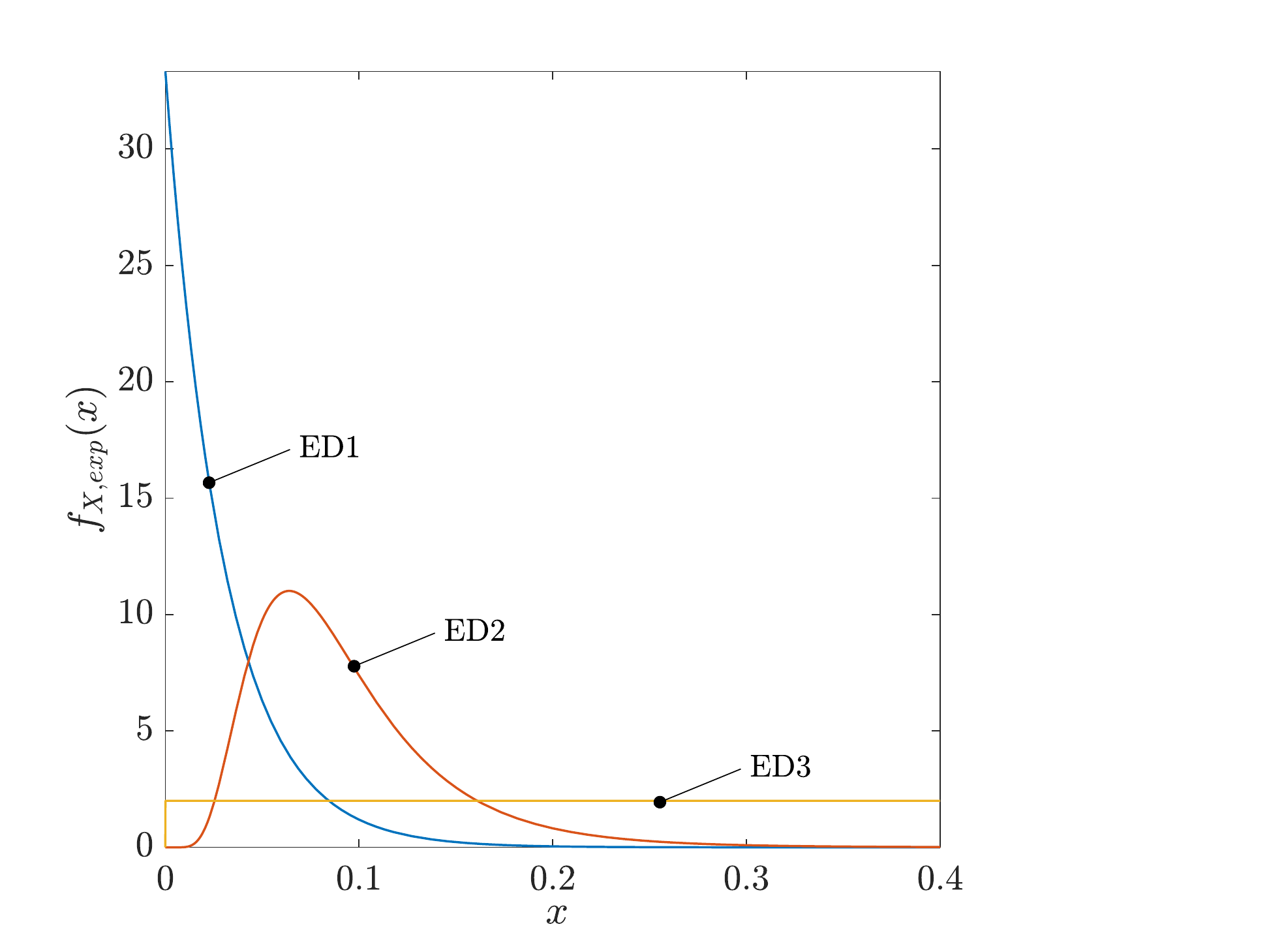}
	\subcaption*{}
\end{subfigure}
\begin{subfigure}{0.5\linewidth}
	\centering
	\small Resulting ROC curves\\
	\includegraphics[scale=0.45, trim=50 0 90 20,clip]{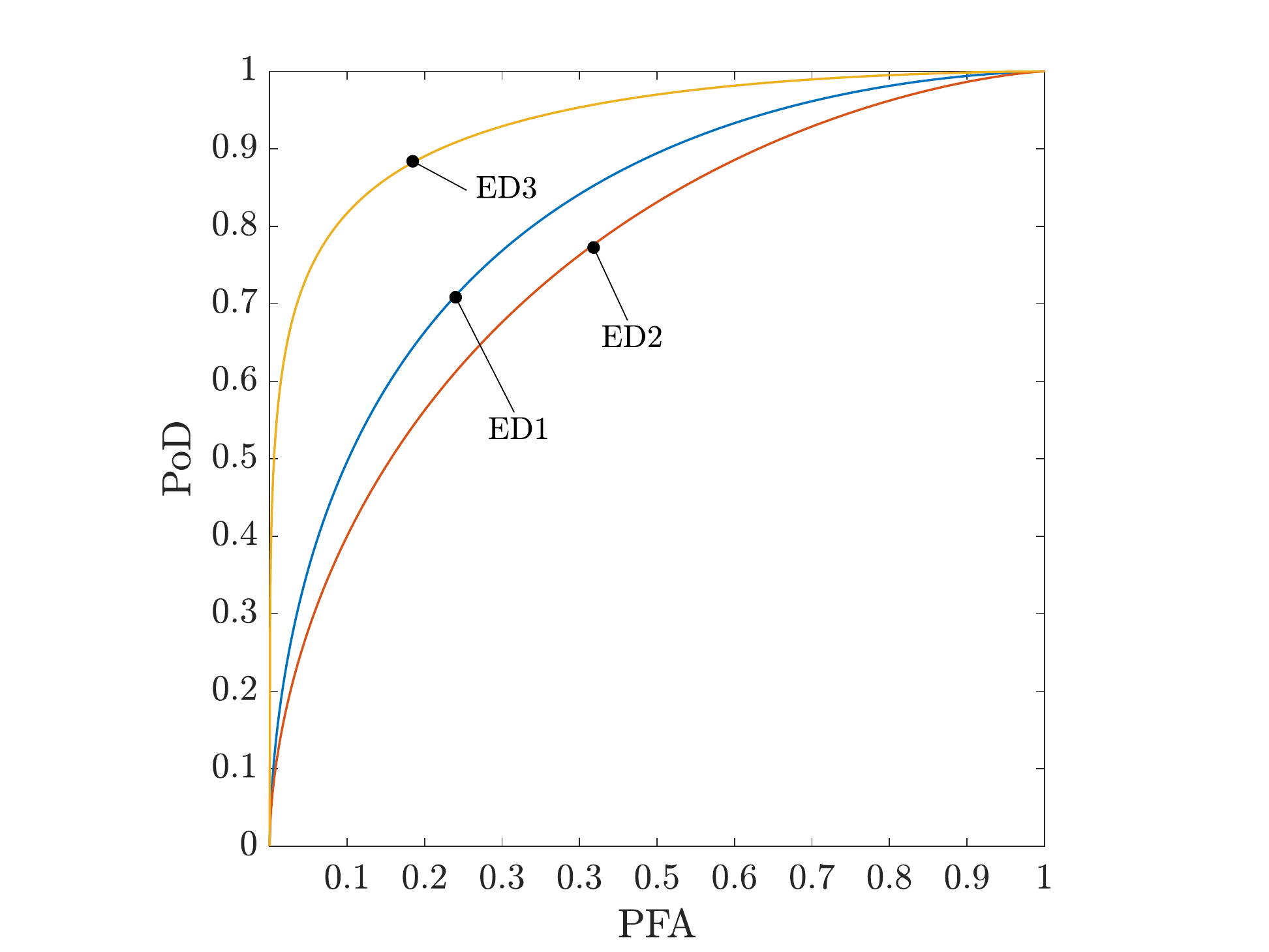}
	\subcaption*{}
\end{subfigure}%
\vspace{-1cm}
\caption{ROC curves for the same NDE derived for different experimental designs, ED1, ED2, and ED3. The model and distributions are presented in the example of Section~\ref{SubS:Example_3and4} and their impact is investigated in Section~\ref{SubS:Ex3_Effect_Exp_Design}. Here, $x_{th}=0.1$.}
\label{Fig:ROC_f_x}
\end{figure}

Eqs.~\eqref{Eq:Mod_1to3_fail} to~\eqref{Eq:PFA} show that the $PoD$ and $PFA$ on the ROC curve are a function of the distribution of the condition $X$. This signifies that even if the ROC curve is evaluated directly from experiments, it is only strictly valid for the distribution of the defects from which it is derived. Therefore, the ROC curve for the same NDE method can vary when applied to different situations. \citet{Boero_et_al_09} illustrates this effect for NDE of deteriorating structures, where the ROC curves change over time along with the distribution of the progressing damage condition. 
A problem appears when the given ROC curve is derived from a previous experiment because the distribution of $X$ in the experiment, which we call \emph{experimental design}, might not match the distribution of $X$ in the specific application. To formally represent this, we distinguish between the pdf $f_X(x)$ and associated cdf $F_X(x)$ of the condition $X$ for a specific application, and the pdf $f_{X,exp}(x)$ and associated cdf $F_{X,exp}(x)$ of $X$ for the experiment from which the ROC curve is derived. Although mentioned in earlier studies \citep{Berens_Hovey_81}, the effect of the experimental design has largely been ignored in the recent literature. To show this effect, we consider three experimental designs and show that the resulting ROC curves change significantly in Fig.~\ref{Fig:ROC_f_x}. An ROC curve from an NDE system provider should therefore be associated with information of the underlying experimental design.


\begin{comment}
\begin{equation}
	PFA(s_{th})=\Pr(S>s_{th}|Y=0)=\int_{s_{th}}^{\infty} f_{S|Y=0}(s)\mathrm{d}s=1-F_{S|Y=0}(s_{th}),
\end{equation}
\begin{equation}\label{Eq:PoDROC}
	PoD(s_{th})=\Pr(S>s_{th}|Y=1)=\int_{s_{th}}^{\infty} f_{S|Y=1}(s)\mathrm{d}s=1-F_{S|Y=1}(s_{th}).
\end{equation}

\end{comment}

\subsection{Model (4): PoD/PFA -- $I$ binary, $Y$ binary} \label{Subs:IY}
This is the most elementary NDE quality model associated with an NDE, which identifies whether the system is in a certain state or not. It is obtained by fixing thresholds both on the condition $X$ and the signal $S$. The associated likelihood is described by a confusion matrix, involving the operating $PFA$ and $PoD$, as presented in  Table~\ref{Tab:Bin_Bin}.
\begin{table}[!h]
\caption{Likelihood of the binary--binary case: The confusion matrix $\Pr(I|Y)$.}
\centering
\renewcommand{\arraystretch}{0.8}
\begin{tabular}{c c c c}
	& &\multicolumn{2}{c}{$\Pr(I|Y)$}\\
	&\multicolumn{1}{c|}{ } &\multicolumn{2}{|l}{\begin{tabular}{c c} $Y=0$& ~~~~~~~~$Y=1$ \end{tabular}}\\
	\hline
	$I$& \multicolumn{1}{c|}{\begin{tabular}{c}	
			0\\ 1 \end{tabular}}  & \begin{tabular}{c}	$1-PFA$ \\ $PFA$ \end{tabular} &\begin{tabular}{c} $1-PoD$\\ $PoD$ \end{tabular}\\	
\end{tabular}
\label{Tab:Bin_Bin}
\end{table}

An example of an NDE system that is described by such a model is ultrasonic flooded member detection, which detects the presence/absence of water in tubular steel members of underwater support structures as an indication of through-thickness cracks or other severe defects \citep{Hayward_et_al_93,Visser_18}. The performance of inspectors for visual inspection or using NDE devices is also typically represented through \replaced{the PoD/PFA model}{ROC curves} \citep{Swets_92,Sheils_et_al_10,daSilva_dePadua_12,Quirk_et_al_18}.

The transition between Model (3) to Model (4) corresponds to calibrating the NDE system to an operating point on an ROC curve, by fixing the threshold $s_{th}$. The $PoD$ and $PFA$ are obtained from Eqs.~\eqref{Eq:PoD_simple} and~\eqref{Eq:PFA_simple}. 

The transition from Model (2) to (4) is obtained by combining Eqs.~\eqref{Eq:PoD} and~\eqref{Eq:PFA} with Eq.~\eqref{Eq:Model_1_to_2}. The $PoD$ and $PFA$ for Model (4) are

\begin{align}\label{Eq:PoD_from2}
PoD=\Pr(I=1|X>x_{th})=\frac{1}{1-F_{X}(x_{th})}\cdot\int_{x_{th}}^{+\infty}{PoD(x)f_{X}(x)\mathrm{d}x},\\\label{Eq:PFA_from2}
PFA=\Pr(I=1|X<x_{th})=\frac{1}{F_{X}(x_{th})}\cdot\int_{-\infty}^{x_{th}}{PoD(x)f_{X}(x)\mathrm{d}x}.
\end{align}

The transition from Model (1) to (4) is obtained by transitioning from Model (1) to (3), then from Model (3) to (4), by fixing thresholds $s_{th}$ and $x_{th}$. Depending on the nature of the condition and the observed signal, it is possible that both thresholds $s_{th}$ and $x_{th}$ are assigned the same value \citep{Sheils_et_al_12}, however they play very distinct roles in the NDE system and should not be confused with one another.



\subsection{Some comments on learning the models}\label{Subs:Model_Uncertainty}
In the ideal case, one would learn the continuous/continuous Model (1) directly, giving a probabilistic relationship between condition $X$ and signal $S$. From this base model, the three other model categories could be derived for specific applications. 
Surprisingly, review of the existing literature reveals that, when continuous/continuous probabilistic models are obtained from experimental or simulated data (see \Cref{Subs:SX}), they are often not used for reliability analysis, or for explicitly deriving PoD curves or other models. 

\begin{figure}[!h]
\centering
\includegraphics[width=1\linewidth]{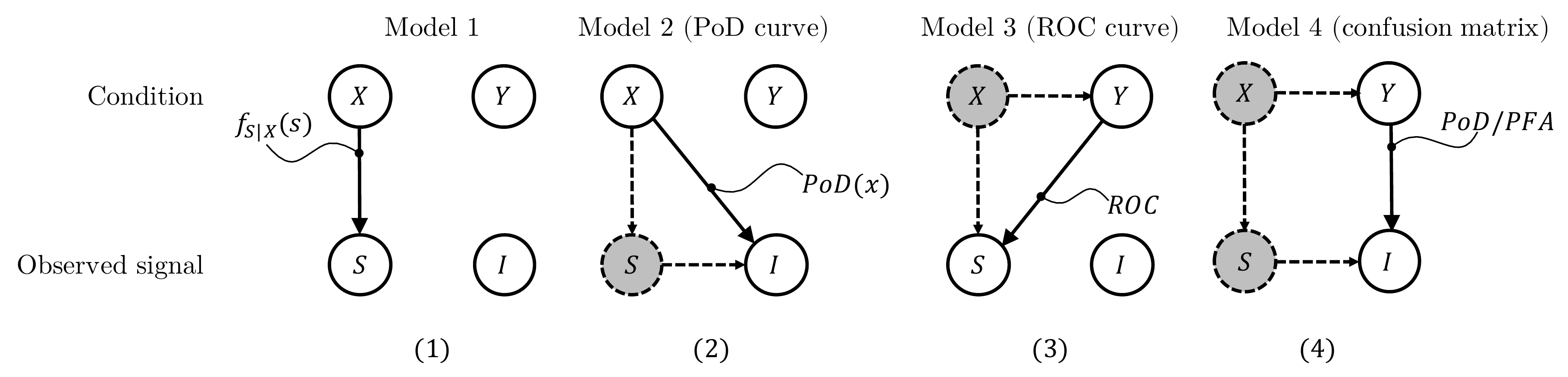}
\caption{Interaction of the four continuous or binary variables $X$, $Y$, $S$, $I$ through the NDE quality Models (1--4). The continuous edges illustrate the relationships between condition and observed signal described by the models. The grayed-out nodes designate the variables hidden when the models are directly evaluated from the experiments. For each model, the link between these hidden variable and the variables of the model can be expressed with Model (1) and an imposed threshold $s_{th}$ or $x_{th}$.}
\label{Fig:Link_Models}
\end{figure}

Fig.~\ref{Fig:Link_Models} summarizes how the four continuous or binary variables $X$, $Y$, $S$, and $I$ interact through the four model types.

It is not always possible to reveal a continuous condition or a continuous signal, in which case the NDE system is described by one of Models (2) to (4). Still, the base model linking $X$ to $S$ can be considered at an abstract level to ensure correct interpretation of the signal and good experimental design.



The correct interpretation of the observed signal is also affected by the uncertainties associated with the model of NDE considered. Several studies point to the fact that experimental data and NDE performance obtained from this experimental data do not translate into similar performance once in-situ \citep[e.g.,][]{Rouhan_Schoefs_03}. The reasons given typically involve aleatoric and epistemic uncertainties in influential parameters affecting the NDE quality, which cannot always be reproduced in an experimental context \citep{Wall_Wedgwood_94,Straub_04}. For instance, temperature, humidity or lighting can affect the performance of an NDE technique \citep{MIL-HDBK-1823A_09} and introduce a bias in the model. \cite{Boero_et_al_09} investigated the spatial variability of NDE models. These model uncertainties can be partly quantified by assigning a probability distribution to the model parameters, thus explicitly accounting for the model uncertainty in the NDE quality model performance \citep{Straub_04}. This hierarchical dependence in the NDE models can also be used to update the models and their parameters with the information collected through Bayesian inference \citep{Aldrin_et_al_13,Hamida_Goulet_20}.

However, even if model uncertainties can be mitigated, the derivations in Sections~\ref{Subs:SY} and~\ref{Subs:IY} show that incorrect assumptions on the distribution of the true condition $X$ lead to erroneous ROC curves and misinterpretation of the performance of NDE systems. In Section~\ref{SubS:Ex3_Effect_Exp_Design} we investigate the effect of such errors on the optimality of decisions.

\section{NDE systems and decision analysis}\label{Sec:DecisionAnalysis}
\subsection{Possibility of improved decision-making}
Section~\ref{Sec:Theory} shows that several choices are made when modeling the performance of an NDE system. There is no unique description of NDE quality and the quality models depend on threshold values on the measured signal and on the condition, $s_{th}$ and $x_{th}$. These thresholds are set by either the designer or the user of the NDE system, often according to standard practice and predefined requirements \citep{Wood_Engle_79,MIL-HDBK-1823A_09,Kurz_et_al_12}. When the decision context is specifically considered for the choice of the thresholds, it is usually through the minimization of mis-classification rates or through minimum requirements on the detection performance of the NDE method \citep{Rajesh_93}. In some studies, the consequences of actions or failure are accounted for \citep[e.g.,][]{Swets_92}. In general, it is acknowledged that the thresholds affect the quality of the NDE system and that they should be calibrated \citep{Kurz_et_al_12,Webb_et_al_15}. 
Furthermore, the repair decision is often directly conditioned on the NDE outcome (recall the name "decision threshold" for $s_{th}$), and the quality of NDE systems is usually assessed without optimizing this decision \citep{Berens_Hovey_81,Sheils_et_al_10}. In this paper, we distinguish between the interpretation of observed signal $S$ (or calibration of the NDE system) and the decision and scrutinize these choices with a formal analysis. 

This section employs formal decision analysis and the value of information (VoI) concept to investigate how NDE quality models affect the decisions taken based on NDE results. As is shown in the numerical results, employing Models (2) to (4) can lead to suboptimal decisions relative to \added{using} Model (1).
To evaluate the performance of NDE systems with Bayesian decision analysis, we introduce the following:

\begin{itemize}
\item[--] $F$ is the critical condition, e.g., a system failure. 
\item[--] The decision maker can choose among a set of actions $\{a_0,a_1,...\}$, typically to mitigate the consequences of failure. The "do nothing" action is denoted by $a_0$.
\item[--] All actions $a_i$ have an associated cost; we assume here that action $a_0$ "do nothing" has an associated cost of $0$. The consequence of $F$ is $c_F$. To be directly comparable, all costs and consequences are expressed in monetary terms.
\end{itemize}

\subsection{Bayesian analysis}\label{SubS:BayesAnalysis}
As stated in Section~\ref{SubS:Mon_Model}, the probabilistic NDE quality models are likelihood functions. When a probabilistic model of the system state is available, Bayesian analysis can be performed, and the reliability of the structure can be evaluated in light of the NDE results. Such reliability updating with likelihoods from NDE quality models was first investigated by \cite{Tang_73}, and many studies integrating reliability analysis with inspection models have been published \citep[e.g.,][]{Madsen_87,Sindel_Rackwitz_96,Onoufriou_Frangopol_02,Straub_11}.

With Bayesian analysis, the posterior probability distribution of the condition $\bm{\Theta}$ is \deleted{as}

\begin{equation}\label{Eq:Posterior_condition}
p(\bm{\theta}|z)\propto \mathcal{L}(\bm{\theta};z)p(\bm{\theta}),
\end{equation}
where $\bm{\Theta}$ is either $X$ or $Y$ depending on the setting. $z$ is the measurement, which is either $s$ or $i$. $\mathcal{L}(\bm{\theta};z)$ is the likelihood function, i.e., one of the NDE quality models of Section~\ref{SubS:Mon_type}. 
The normalizing constant of Eq.~\eqref{Eq:Posterior_condition} is the model evidence, $p(z)$:
\begin{equation}\label{Eq:Mod_evidence}
p(z)=\mathbf{E}_{\bm{\Theta}}\left[\mathcal{L}(\bm{\theta};z)\right],
\end{equation}
where $\mathbf{E}_{\bm{\Theta}}[\cdot]$ is the expectation with respect to $p(\bm{\theta})$.

One is typically interested in identifying a critical condition, or failure. Conditional on the NDE outcomes, one obtains the posterior probability of this critical condition $\Pr(F|Z=z)$ as
\begin{equation}\label{Eq:Posterior_failure}
\Pr(F|Z=z)=\frac{1}{p(z)}\mathbf{E}_{\bm{\Theta}}\left[\Pr(F|\bm{\theta})\mathcal{L}(\bm{\theta};z)\right].
\end{equation}


\subsection{Optimal decision and value of information}\label{SubS:VoI}
While the methods developed in the NDE community focus on comparing inspection techniques, the actual value added is related to how the information from NDE leads to better decisions \citep{Raiffa_Schlaifer_61,Straub_14}. In decision analysis, it is assumed that the decision maker selects an action that maximizes the expected utility after obtaining information $Z$ through NDE. Here we consider utility to be negatively proportional to costs. Hence, the optimization problem is written in as
\begin{equation}\label{Eq:Elem_opt}
a_{opt}(z)=\arg \min_{a\in\{a_0,a_1...\}} \mathbf{E}_{\bm{\Theta}|z}[C_T(a,\bm{\Theta})].
\end{equation}
The total cost $C_T$ includes the cost of the actions and consequences of failure. $\mathbf{E}_{\bm{\Theta}|z}[\cdot]$ is the \deleted{conditional} expectation with respect to the conditional distribution $p(\bm{\theta}|z)$ from Eq.~\eqref{Eq:Posterior_condition}. 

\begin{figure}[htpb]
\centering
\includegraphics[width=0.5\linewidth]{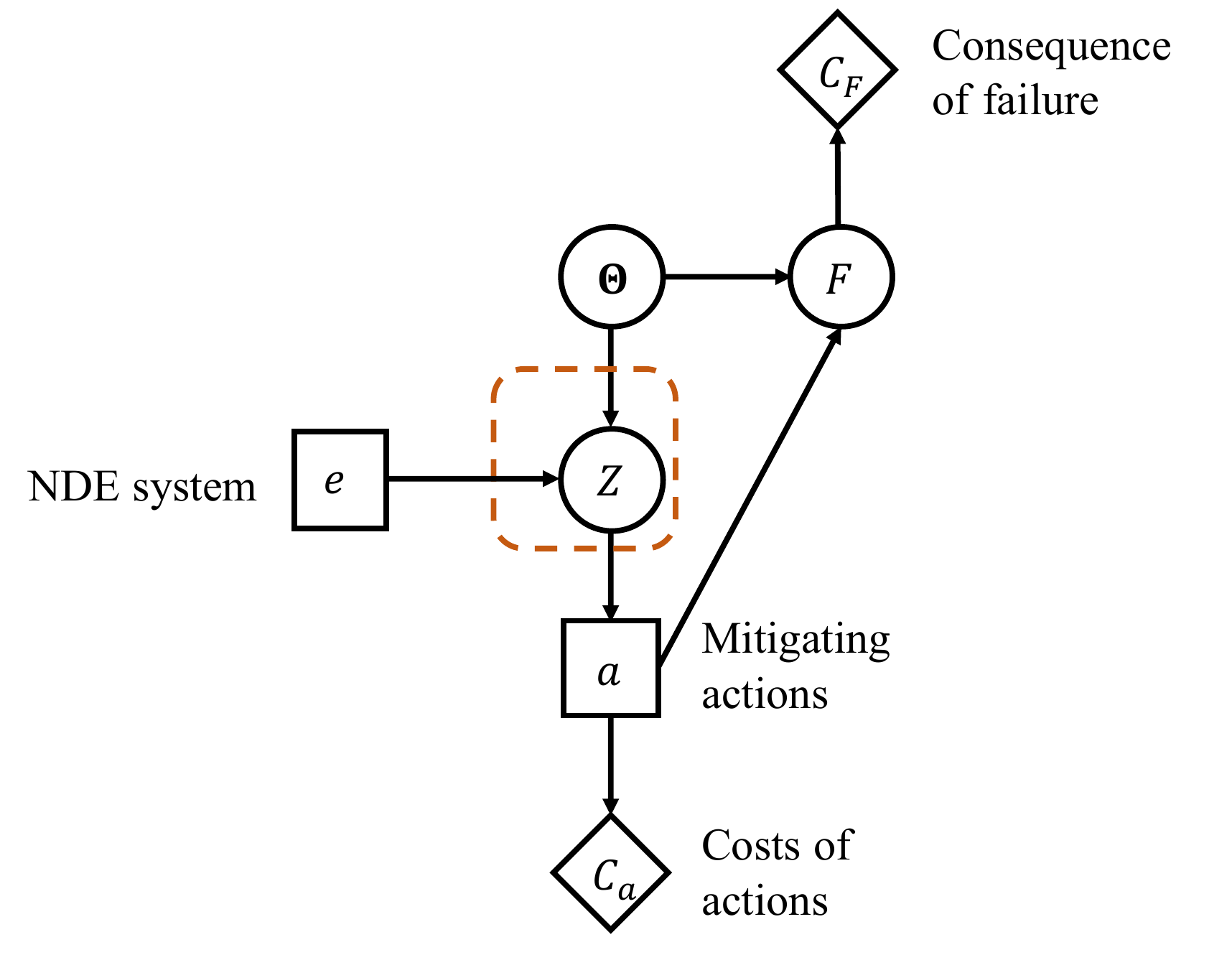}
\caption{Influence diagram for the basic decision problem. The true condition, the observation, and the failure event are represented by round nodes, the actions and choice of NDE system by square nodes, and the costs incurred as diamond nodes.}
\label{Fig:ID_simple}
\end{figure}

The influence diagram of Fig.~\ref{Fig:ID_simple} represents the probabilistic relationships between the true condition, the observation and the failure event (round nodes), the actions and choice of NDE system (square nodes), as well as the costs incurred (diamond nodes).
For example, $e$ can be a visual inspection of a crack, the outcome $Z$ can be the signal $I$ "detection" or "no detection," and the possible actions $a$ "repair" or "no repair." 

The observation $Z$ is \deleted{a priori} unknown \added{a priori} and is described by the distribution of Eq.~\eqref{Eq:Mod_evidence}. Hence, the expected cost resulting from implementing the NDE system $e$, $C_e$, is obtained by deriving an expected value with respect to $p(z)$, i.e.,
\begin{equation}\label{Eq:Exp_cost}
C_e=\mathbf{E}_{Z}\left[\mathbf{E}_{\bm{\Theta}|Z}[C_T(a_{opt}(Z),\bm{\theta})]\right].
\end{equation}

\cite{Raiffa_Schlaifer_61} denote with $e_0$ the case where no information is collected and the decisions are based on the prior knowledge of the system. The optimal action and expected cost under $e_0$ are

\begin{align}
\label{Eq:Opt_e0}a_{e_0}=\arg \min_{a} \mathbf{E}_{\bm{\Theta}}[C_T(a,\bm{\Theta})],\\
\label{Eq:Exp_cost_prior}C_0=\mathbf{E}_{\bm{\Theta}}[C_T(a_{e_0},\bm{\Theta})].
\end{align}

The value of information (VoI) (also called expected value of sample information) for an NDE system $e$ is the difference between the total expected prior cost associated with $e_0$ and the total expected cost associated with $e$.
\begin{equation}\label{Eq:VoI}
VoI(e)=C_0-C_e.
\end{equation}

Notably, if for any outcome $z$ obtained through $e$, the optimal action $a_{opt}(z)=a_{e_0}$, then $VoI(e)=0$. In other terms, an NDE system only brings value if it changes the decision on which action to take. The VoI does not include the cost of implementing the NDE system $e$ (e.g., cost of mobilizing a team of inspectors). This cost can be added to  $C_e$ in Eq.~\eqref{Eq:VoI} to obtain the net VoI (NVoI). 
In the Bayesian decision analysis framework, an NDE system with a positive NVoI is worth implementing, and the higher the NVoI the better. \cite{Somoza_et_al_90} employed this concept to calibrate an NDE system for optimal maintenance planning.

Although the concept of VoI appeared in the mid-20th century, it was not used to support inspection and maintenance decisions until recently, mainly because of the complexity associated with its calculation. Reliance on established processes also explains why it was left aside. Historically, the primary focus of NDE performance evaluation has been to support systematic inspection plans or condition-based maintenance, where monitoring outcomes systematically result in one action, of the "find nothing, do nothing" and "find a defect then repair" type \citep{Berens_Hovey_81,KTA_16}.  The performances of NDE techniques are codified by standards, based on traditional rule-of-thumb criteria, that are not specific to the problem. 
An example of NDE for systematic maintenance is the automated Eddy current inspection system for detecting and repairing cracks in engine components which has been used by the US air force since the 1980s \citep{Berens_Vol1_00}. In the nuclear industry, standards prescribe fixed time intervals between inspections and the NDE techniques to adopt, assuming that any detected defect is systematically repaired \citep{KTA_16}. The conservative conditions for inspection and repair are adopted without a risk-based decision analysis. In this context, false alarms and unnecessary repairs are judged as negligible when compared to preserving the safe operation of the system.

Currently, NDE is used for integrity management and service life extension of large infrastructure such as bridges. The cost of unnecessary repair is not negligible and can vary greatly with the accessibility of the structure. The VoI allows accounting for the economic aspects in the decision process. The computational and theoretical advances in the computation of the VoI in recent years provide the opportunity to evaluate the performance of an NDE system in the real decision-making context.

In Sections~\ref{SubS:Example_3and4} and~\ref{SubS:Example_1and2} we perform optimizations using Eqs.~\eqref{Eq:Elem_opt} to~\eqref{Eq:VoI}, where the decision maker can decide whether or not to repair the system based on observations provided by an NDE system. In the one-step decision problem, Eq.~\eqref{Eq:Elem_opt} reduces to a simple condition that links the likelihood $\mathcal{L}(\bm{\theta}; z)$ with the different costs and prior probabilities. 


\section{Example 1: Hypothetical NDE system}\label{SubS:Example_3and4}
\added{The purpose of this example is to demonstrate the unifying framework \deleted{on a purely theoretical NDE system} and investigate the impact of each of the four NDE models on the optimal decision outcome and cost in a basic decision problem. We highlight the effect of experimental design on the assessment of NDE performance.}
\subsection{Base model}\label{SubS:Base_hypothetical}
We consider a hypothetical NDE system which measures a continuous condition and outputs a continuous signal, in analogy with crack detection systems such as UT.

The conditional pdf of the continuous measurement signal $S$ given true condition $X$ is
\begin{equation}\label{Eq:Ex_3_full_pdf}
f_{S|X}(s|x)=\frac{1}{s\sqrt{2\pi}}\exp\left(-\frac{\left(\ln s-\ln\left(2x^3+x^2+10^{-2}\exp(-1/2)\right)\right)^2}{2}\right).
\end{equation}

This is the lognormal distribution with parameters $[\ln\left(2X^3+X^2+10^{-2}\exp(-1/2)\right),1]$. This conditional probability density is shown in Fig.~\ref{Fig:unified_model} (Model (1)).
\begin{comment}
\begin{figure}[!h]
\centering
\includegraphics[scale=0.5]{Conditional_f_s_x_const_var}
\caption{Conditional probability density $f_{S|X}$.}
\label{Fig:Ex_3_cond_density}
\end{figure}
\end{comment}

In the following we consider a basic decision problem, where one needs to decide on a repair action. We adopt the four different NDE quality models described in Section~\ref{SubS:Mon_type}  in turn and assess their impact on the repair decision. We additionally investigate the effect of learning the NDE models with different experimental designs.

\subsection{Solution of the one-step decision problem}\label{App:Sol_one_step}
The decision problem consists in selecting a repair action to mitigate the consequence of failure, after obtaining imperfect NDE outcome $Z$ on state $\bm{\Theta}$. It is illustrated in Fig.~\ref{Fig:ID_simple}. The condition $\bm{\Theta}$ is either $X$ or $Y$. The observation variable is $Z$, which is either the continuous signal $S$ or the binary signal $I$. 
$\mathcal{L}(\bm{\Theta};z)$ indicates the likelihood function.
In this setup, we consider two possible actions: either do nothing $a_0$, for a cost $c_A(a_0)=0$, or repair $a_R$, for a cost $c_A(a_R)=c_R$. The probability of system failure $F$ is defined conditional on the state $\bm{\Theta}$ and the action $A$. The consequence of failure is $c_F$.

The a priori optimal action $a_{e_0}$ is the one that minimizes the total expected cost as given by Eq.~\eqref{Eq:Exp_cost_prior}. Here,

\begin{align}
\mathbf{E}_{\bm{\Theta}}[C_T(a,\bm{\Theta})]=c_A(a)+c_F\Pr(F|a) ~ \text{with} ~ 
\Pr(F|a)=\mathbf{E}_{\bm{\Theta}}\left[\Pr(F|\bm{\Theta},a)\right].
\end{align}
From Eq.~\eqref{Eq:Opt_e0}, one finds that

\begin{align}\label{Eq:OneStep_prior}
a_{e_0}=a_R \iff c_F\Pr(F|a_0)>c_R+c_F\Pr(F|a_R).
\end{align}
As per Eq.~\eqref{Eq:Elem_opt}, the a posteriori optimal action is the one that minimizes the conditional expected total cost, 

\begin{align}
\label{Eq:Exp_Cost_action}\mathbf{E}_{\bm{\Theta}|z}[C_T(a,\bm{\Theta})]=c_A(a)+c_F\Pr(F|Z=z,a),~\text{with}~
\Pr(F|Z=z,a)=\mathbf{E}_{\bm{\Theta}|z}\left[\Pr(F|\bm{\Theta},a)\right].
\end{align}

Making the likelihood $\mathcal{L}(\bm{\Theta};z)$ explicit in the conditional expectation, one obtains that

\begin{align}
\label{Eq:Sol_oneStep_cont_Cond} a_{opt}(z)=a_R\iff  \mathbf{E}_{\bm{\Theta}}\left[\mathcal{L}(\bm{\Theta};z) c_F\Pr(F|\bm{\Theta},a_0)\right]>\mathbf{E}_{\bm{\Theta}}\left[\mathcal{L}(\bm{\Theta};z)\left\{ c_R+c_F\Pr(F|\bm{\Theta},a_R) \right\}\right].
\end{align}
When the condition is binary, i.e., $\bm{\Theta}=Y$, $\mathcal{L}(Y;z)$ is the likelihood for Models (3) or (4). The condition of the above equation can be transformed into a condition on $\frac{\mathcal{L}(Y=1;z)}{\mathcal{L}(Y=0;z)}$, also called the "likelihood ratio test" \citep{Peterson_Birdsall_53,Green_Swets_66}. If $a_{e_0}=a_R$,

\begin{align}\label{Eq:Sol_oneStep_Bin_Cond}
a_{opt}(z)=a_R&\iff\frac{\mathcal{L}(Y=1;z)}{\mathcal{L}(Y=0;z)}>\frac{\left\lbrace c_R+c_F\Pr(F|Y=0,a_R)-c_F\Pr(F|Y=0,a_0)\right\rbrace\Pr(Y=0)}{\left\lbrace c_F\Pr(F|Y=1,a_0)-\left[{c_R+c_F\Pr(F|Y=1,a_R)}\right]\right\rbrace\Pr(Y=1)}.
\end{align}
The expected total cost associated with the optimal action following an NDE is given by Eq.~\eqref{Eq:Exp_cost}.
When the observation is binary, i.e., $Z=I$ and $\{I=1\}=\{S>s_{th}\}$, the expected cost depends on the choice of threshold $s_{th}$.


\subsection{Condition model and a priori optimal action}

The prior pdf of $X$ is the exponential pdf with mean $0.03$.
\begin{equation}\label{Eq:True_dist}
f_{X}(x)=\frac{1}{0.03}\exp\left(-\frac{x}{0.03}\right).
\end{equation}

\begin{figure}[!h]
\centering
\includegraphics[scale=0.5]{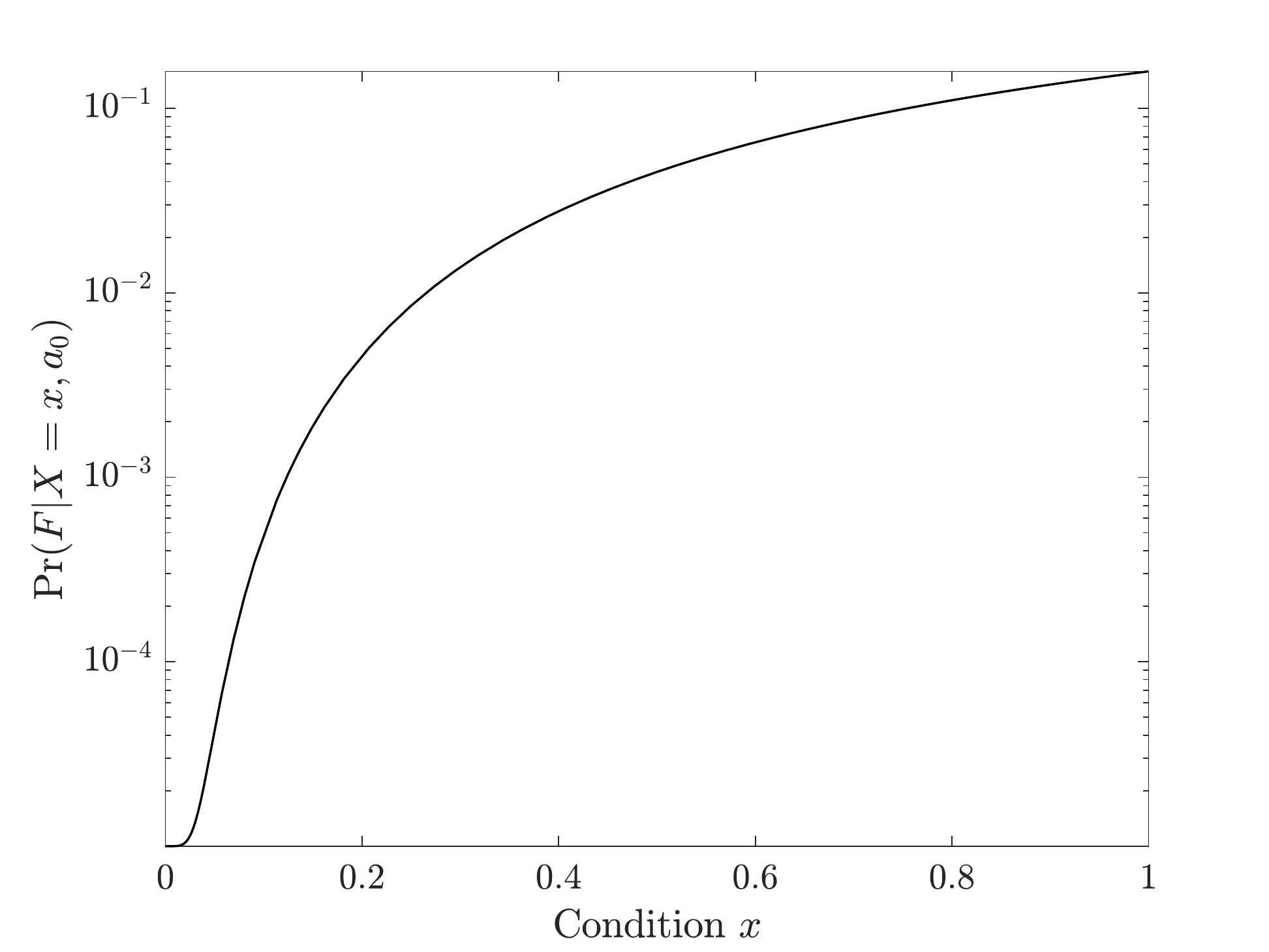}
\caption{Probability of system failure conditional on the condition $X$.}
\label{Fig:PrFX}
\end{figure}

We fix the cost of repair $c_R=1$. 
Under the do-nothing action $a_0$, the probability of failure of the system is defined conditional on $X$ (Fig.~\ref{Fig:PrFX}),
\begin{equation}
\Pr(F|X=x)=10^{-5}+(1-10^{-5})\cdot \left(\frac{1}{2}+\frac{1}{2}\text{erf}\left(\frac{\log x-0.1}{\sqrt{2}}\right)\right),
\end{equation}
where $\text{erf}$ is the error function.

When the system is repaired, the probability of failure is reduced such that $\forall x,~ \Pr(F|X=x,a_R)=p_{F|R}=10^{-4}$. The cost of failure is $c_F=800$.


Eq.~\eqref{Eq:OneStep_prior} gives the condition to identify the a priori best decision. The probability of failure of the system conditional on action $a_0$ is evaluated as $\Pr(F|a_0)=1.2\cdot 10^{-3}$, and the probability of failure conditional on action $a_R$ is $\Pr(F|a_R)=p_{F|R}=10^{-4}$. Since $c_F\Pr(F|a_0)=0.94$ is smaller than $c_R+c_Fp_{F|R}=1.08$, the a priori best decision is therefore to do nothing, $a_{e_0}=a_0$, with associated cost $0.94$.

\subsection{A posteriori optimal actions with Model (1)}
\begin{figure}[!h]
\centering
\includegraphics[width=0.7\linewidth, clip, trim=30 0 30 0]{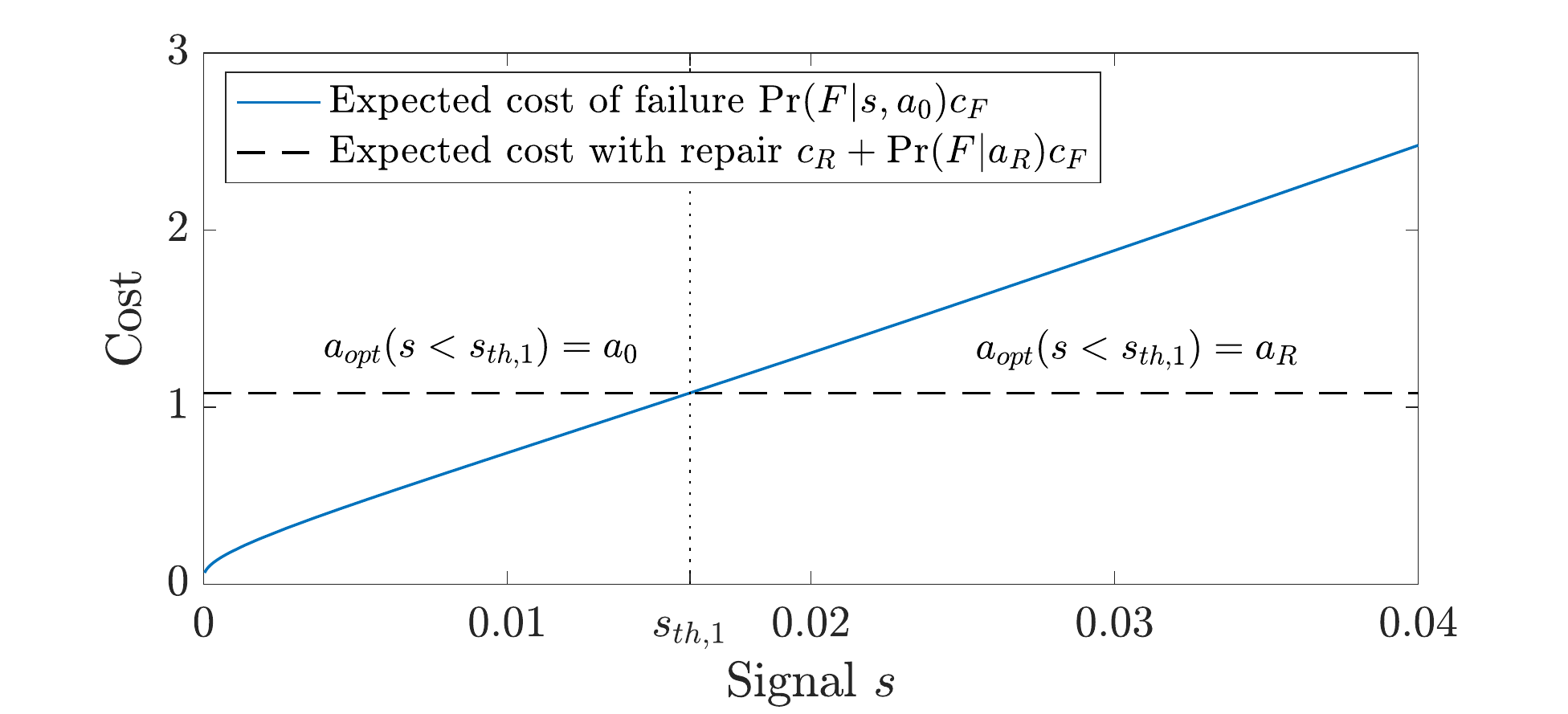}
\caption{Expected cost for action $a_0$ and $a_R$ conditional on the observed signal $S$.}
\label{Fig:Ex_3_CondExp_cost_PoD}
\end{figure}

The likelihood $\mathcal{L}(x;s)$ is given by Eq.~\eqref{Eq:Ex_3_full_pdf}. By solving Eq.~\eqref{Eq:Sol_oneStep_cont_Cond} for $z=s$, we find that the optimal action is 
$a_{opt}(s)=a_0$ for $s<s_{th,1}=1.6\cdot10^{-2}$, and $a_{opt}(s)=a_R$ for  $s>s_{th,1}$. \added{Here, $s_{th,1}$ is effectively a repair threshold}. The comparison of expected costs for both actions as a function of the measured signal $S$ is depicted in Fig.~\ref{Fig:Ex_3_CondExp_cost_PoD}.
The expected total cost computed with Eq.~\eqref{Eq:Exp_cost} is $0.65$. The VoI of this NDE system is $0.94-0.65=0.29$.


\subsection{A posteriori optimal actions for a given PoD curve}
\begin{figure}[!h]
\centering
\includegraphics[width=0.7\linewidth]{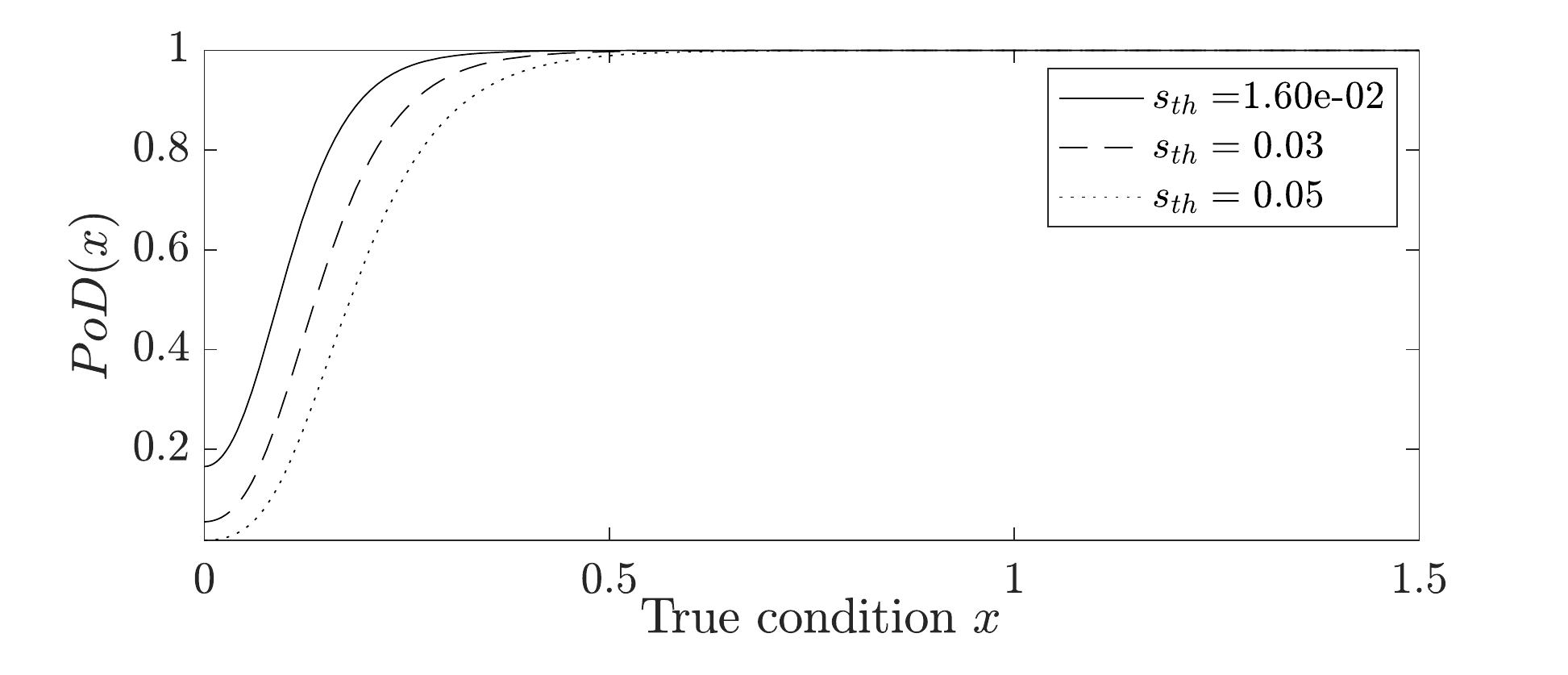}
\caption{PoD curves for different thresholds $s_{th}$, including $s_{th,2}$.}
\label{Fig:Ex_3_OptPoD}
\end{figure}

We now determine the optimal a posteriori action identified with the PoD curve (Model (2)) and the associated VoI. We remind that in Model (2), a binary signal $I$ is considered, which is related to the continuous signal $S$ by $\{I=1\}=\{S>s_{th}\}$. The likelihood $\mathcal{L}(x;I)$ is given by the PoD curve of Eq.~\eqref{Eq:Model_1_to_2}, which depends on the fixed threshold $s_{th}$. PoD curves for different thresholds are depicted in Fig.~\ref{Fig:Ex_3_OptPoD}.

PoD curves are typically given and not specific to the application, and hence are likely suboptimal for a given decision context. For example, if the given PoD curve is calibrated with $s_{th}=0.03$, the optimal actions are $a_{opt}(I=0)=a_0$, $a_{opt}(I=1)=a_R$ and the expected cost is $0.70$. The resulting the VoI is $0.94-0.70=0.24$, which is below the potentially achievable VoI$=0.29$. 

\begin{figure}[!h]
\centering
\includegraphics[width=0.7\linewidth]{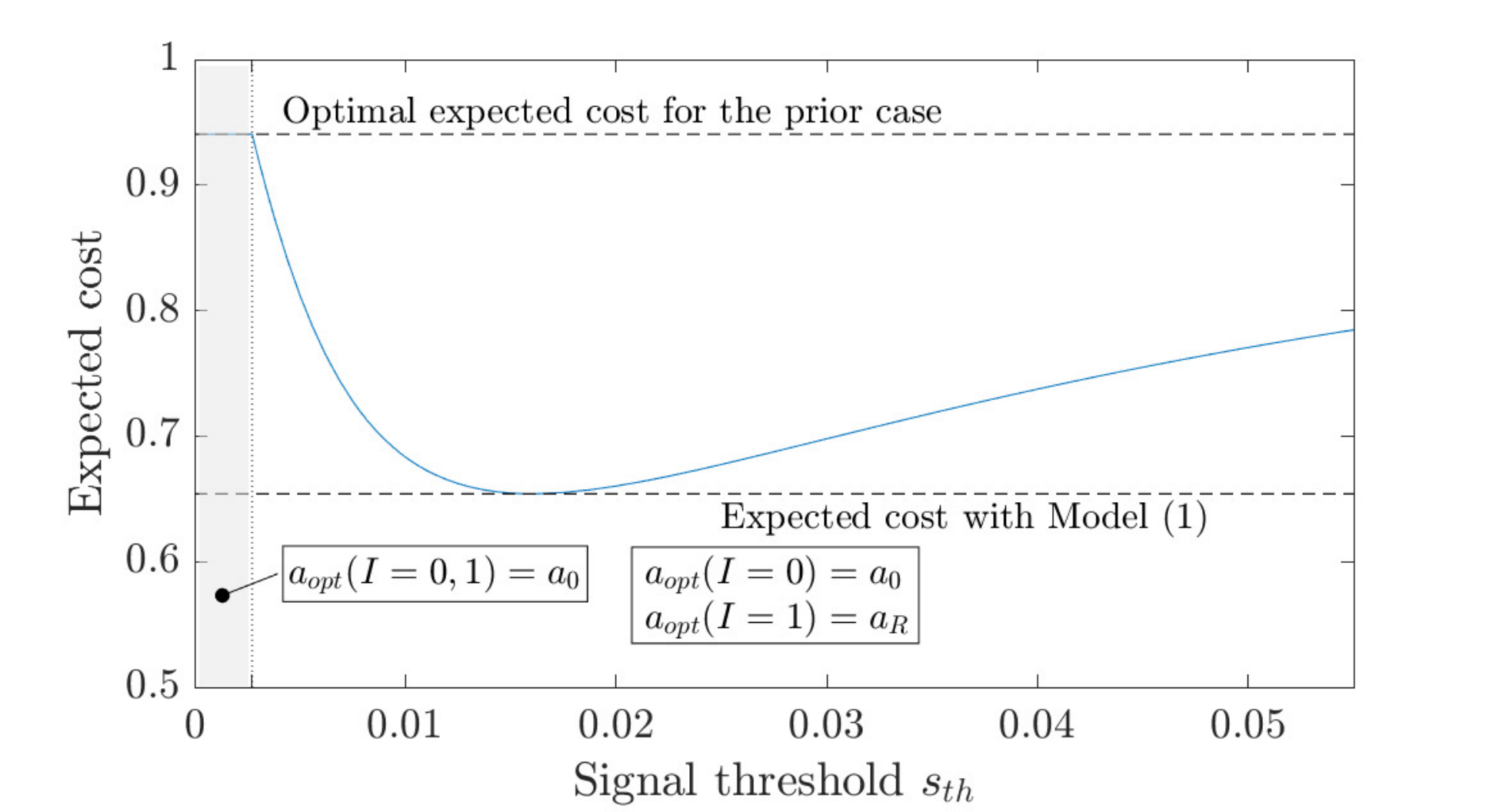}
\caption{Expected cost using the PoD curve model as a function of the fixed signal threshold $s_{th}$.} 
\label{Fig:Ex_3_Exp_cost_PoD}
\end{figure}

For each imposed threshold $s_{th}$ and associated PoD curve, Fig.~\ref{Fig:Ex_3_Exp_cost_PoD} indicates the optimal actions $a_{opt}(I)$ and the expected cost calculated with Eq.~\eqref{Eq:Exp_cost}. A threshold exists that maximizes the VoI and minimizes the expected cost for this case study. It is the decision threshold $1.6\cdot10^{-2}$, with the associated expected cost of $0.65$, which corresponds to the result obtained with Model (1). However, when describing the NDE by the PoD curve, this optimal solution will only be obtained by coincidence. In general, the PoD curve leads to a suboptimal decision.

\subsection{A posteriori optimal actions for a given ROC curve}\label{SubS:Ex3_ROC}
When ROC curves are utilized to represent the quality of NDE, the continuous condition $X$ is replaced by the binary condition $\{Y=1\}=\{X>x_{th}\}$. Here we fix $x_{th}=0.1$. The corresponding ROC curve is depicted in Fig.~\ref{Fig:ROC_f_x}, where the distribution of $X$ corresponds to the experimental design ED1.

To compare the performance of Models (1) and (3), we must first ensure that the relationships between binary condition $Y$ and failure $F$ are compatible with the relationship between continuous condition $X$ and failure $F$. This compatibility is ensured by the following equations,
\begin{comment}
\begin{align}
&\Pr(Y)=\int_{X}{\Pr(Y|X=x)f_X(x)\mathrm{d}x}\\
&\Pr(F|Y)=\int_{X}{\Pr(F|X=x)f_{X|Y}(x)\mathrm{d}x}=\frac{\int_{X}{\Pr(F|X=x)\Pr(Y|X=x)f_{X}(x)\mathrm{d}x}}{\Pr(Y)}
\end{align}
where $\Pr(Y=1|X=x)=\mathbbm{1}_{x>x_{th}}$.
\end{comment}

\begin{align}
\label{Eq:PrY}&\Pr(Y=1)=1-F_X(x_{th})=3.6\cdot 10^{-2},\\
&\Pr(F|Y=1,a_0)=\frac{\int_{x_{th}}^{+\infty}{\Pr(F|X=x)f_{X}(x)\mathrm{d}x}}{1-F_X(x_{th})}=1.7\cdot 10^{-2},\\
&\Pr(F|Y=0,a_0)=\frac{\int_{0}^{x_{th}}{\Pr(F|X=x)f_{X}(x)\mathrm{d}x}}{F_X(x_{th})}=5.8\cdot 10^{-4}.
\end{align}
Additionally, it is $\forall y\in\{0,1\},~\Pr(F|Y=y,a_R)=p_{F|R}$.

The solution of the decision problem is derived from Eq.~\eqref{Eq:Sol_oneStep_Bin_Cond}:

\begin{align}\label{Eq:Ex_3_ROC_lik_ratio}
a_{opt}(s)=a_R\iff\frac{\mathcal{L}(Y=1;s)}{\mathcal{L}(Y=0;s)}=\frac{f_{S|Y=1}(s)}{f_{S|Y=0}(s)}>1.31.
\end{align}

The likelihoods $\mathcal{L}(Y;s)=f_{{S|Y}}(s)$ are computed with Eqs.~\eqref{Eq:Mod_1to3_fail} and~\eqref{Eq:Mod_1to3_safe}.
Both Models (1) and (3) provide the same model evidence $p(s)$. 

From Eq.~\eqref{Eq:Ex_3_ROC_lik_ratio}, we find the optimal action $a_{opt}(s)$ is $a_0$ when $s<1.7\cdot 10^{-2}$ and $a_R$ otherwise. This threshold $s_{th,3}=1.7\cdot 10^{-2}$ is shown as the optimal operating point on the ROC curve in  Fig~\ref{Fig:Ex3_Exp_cost_ROC}. Furthermore, Fig.~\ref{Fig:Ex3_Exp_cost_ROC} indicates that ROC performance indices, such as the Youden Index, are associated with certain operating points. But using either of these points, as proposed in some references \citep[e.g.,][]{Schoefs_Clement_04}, is not optimal in view of the specific decision. 

The total expected cost for the optimal operating point is $0.76$. 
The VoI of this NDE system is $0.18$, lower than for Model (1). Even though the actions are optimized, their efficiency are limited by the fixed threshold $x_{th}$ and associated "failure" domain of the NDE device, through Eq.~\eqref{Eq:Ex_3_ROC_lik_ratio}.
Other values of $x_{th}$ can yield a higher or lower VoI: E.g., with $x_{th}=0.2$ the associated optimal expected cost is $0.91$ and the VoI is only $0.03$. However, the threshold $x_{th}$ can typically not be influenced for a given NDE device.

\subsection{A posteriori optimal actions for a given point on the ROC curve}\label{SubS:Ex3_PoDPFA}
Here, the NDE system is described by a point on the ROC curve defined above in Section~\ref{SubS:Ex3_ROC}, such that $\{I=1\}=\{S>s_{th}\}$.
The likelihood of Model (4) $\mathcal{L}(Y;I)$ is given by a confusion matrix  as per Table~\ref{Tab:Bin_Bin}, with $\mathcal{L}(Y=1; I=1)=PoD$ and $\mathcal{L}(Y=0; I=1)=PFA$.
\begin{comment}
For a given $PoD$ and $PFA$, the probability of failure conditional on the signal $I=1$ is
\begin{equation}
\Pr(F|I=1)=\frac{1}{\Pr(I=1)}\left(PFA\int_{0}^{x_{th}}\Pr(F|X=x)f_X(x)\mathrm{d}x+PoD\int_{x_{th}}^{+\infty}\Pr(F|X=x)f_X(x)\mathrm{d}x\right)
\end{equation}
where
\begin{equation}
\Pr(I=1)=PFA\cdot F_X(x_{th})+PoD\cdot(1-F_X(x_{th}))
\end{equation}
\end{comment}
As in Eq.~\eqref{Eq:Ex_3_ROC_lik_ratio}, the likelihood ratio verifies the conditions

\begin{align}
&a_{opt}(I=1)=a_R\iff\frac{\mathcal{L}(Y=1;I=1)}{\mathcal{L}(Y=0;I=1)}>1.31\iff\frac{PoD}{PFA}>1.31,\\
&a_{opt}(I=0)=a_R\iff\frac{\mathcal{L}(Y=1;I=0)}{\mathcal{L}(Y=0;I=0)}>1.31\iff\frac{1-PoD}{1-PFA}>1.31.
\end{align}
For example, for $s_{th}=1\cdot10^{-2}$ (and $x_{th}=0.1$ as above), $PoD=0.82$, $PFA=0.37$, $a_{opt}(I=1)=a_R$, $a_{opt}(I=0)=a_0$ and the expected cost is $0.79$.

\begin{figure}[!h]
\centering
\includegraphics[width=0.7\linewidth]{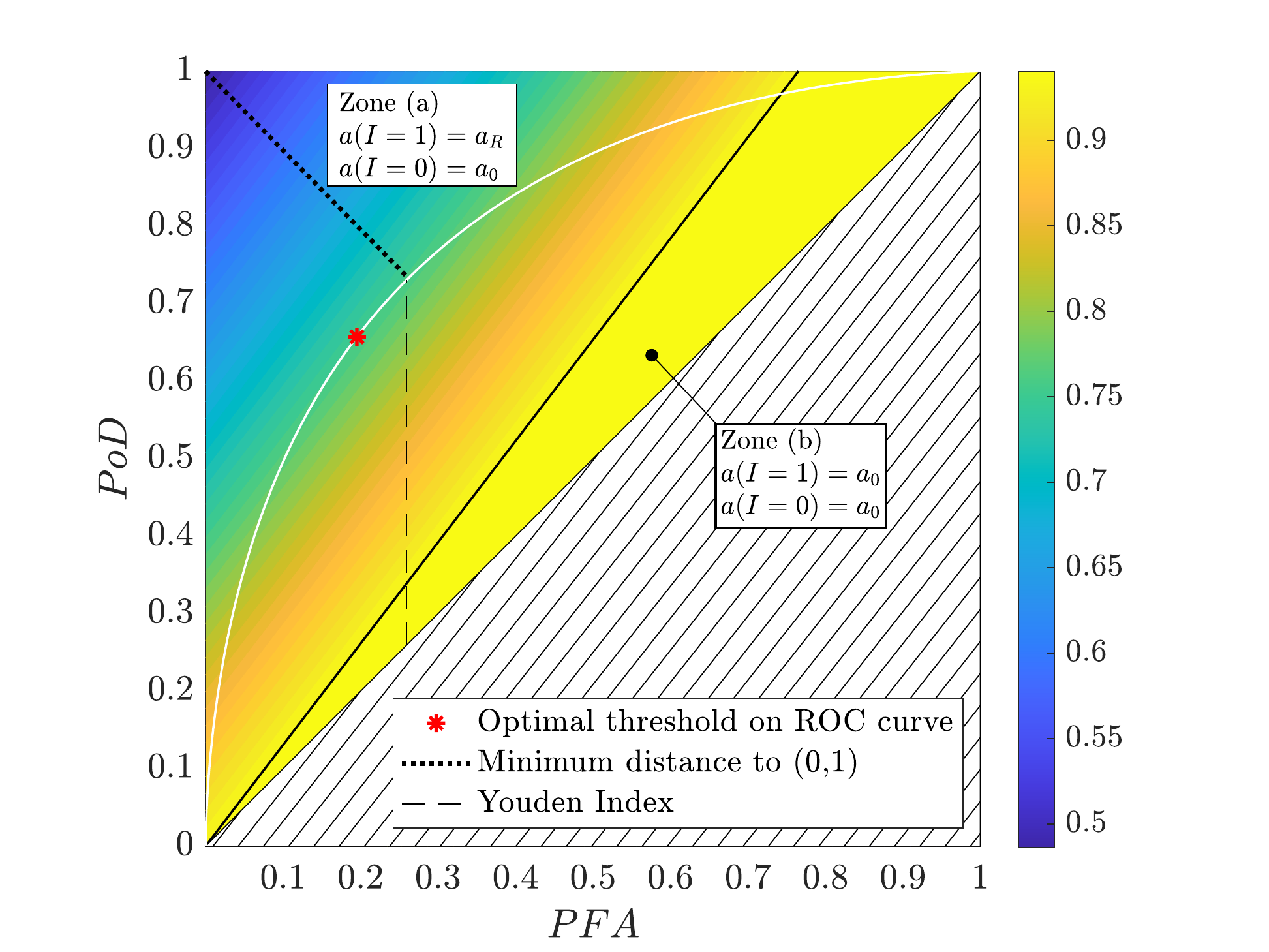}
\caption{Optimal expected cost as a function of $PFA$ and $PoD$. Zone (b) corresponds to $VoI=0$: this means that for any decision threshold $s_{th}$ on the ROC curve that falls within that zone, the best course of action is the prior optimal action, i.e., do nothing. The hatched area corresponds to the situation where the significance of $I=0$ and $I=1$ are inverted.}
\label{Fig:Ex3_Exp_cost_ROC}
\end{figure}

Fig.~\ref{Fig:Ex3_Exp_cost_ROC} shows the expected cost for any fixed $PoD$ and $PFA$ as well as the optimal actions $a_{opt}(I)$. Two zones can be distinguished: (a) for any $(PFA,PoD)$ in the yellow zone, the optimal course of action is to do nothing, whatever the observation outcome; this is what the prior optimal action also prescribes, therefore the VoI at these points is 0; (b) for any $(PFA,PoD)$ to the left of this yellow zone, detection triggers repair $a_R$, and no detection entails $a_0$. 

\subsection{Influence of the experimental design}\label{SubS:Ex3_Effect_Exp_Design}


We investigate how the experimental design $f_{X,exp}$ used to learn the ROC curve can affect the decision taken and the resulting expected total cost and VoI.
We consider three experimental designs ED1, ED2, and ED3, where the test conditions follow an exponential, lognormal, and uniform distribution, respectively (Eqs.~\eqref{Eq:Exp_densities_1} to~\eqref{Eq:Exp_densities_3}). The imposed threshold on the condition remains $x_{th}=0.1$ as above, with $\{Y=1\}=\{X>x_{th}\}$.

\begin{align}\label{Eq:Exp_densities_1}
&f_{X,exp,1}(x)=\frac{1}{0.03}\exp\left(-\frac{x}{0.03}\right),\\
&f_{X,exp,2}(x)=\frac{2}{x\sqrt{2\pi}}\exp\left(-2(\ln x+2.5)^2\right),\\
\label{Eq:Exp_densities_3}
&f_{X,exp,3}(x)=2\text{~if~} 0\leq x\leq 0.5,~ 0 \text{~otherwise.}
\end{align}

The corresponding likelihoods $f_{{S|Y},exp,i}(s)$ are obtained with Eqs.~\eqref{Eq:Mod_1to3_fail} and~\eqref{Eq:Mod_1to3_safe}.
The resulting ROC curves for $x_{th}=0.1$ are plotted in Fig.~\ref{Fig:ROC_f_x}.




For each likelihood $f_{{S|Y},exp,i}(s)$, the model evidence $p_{exp,i}(s)$ is evaluated with Eq.~\eqref{Eq:Mod_evidence}, where the prior probability of the condition $Y$ is obtained with Eq.~\eqref{Eq:PrY}:

\begin{align}
\label{Eq:Mod_Evidence}p_{exp,i}(s)=\sum_{y=0}^{1}{f_{S|Y=y,exp,i}(s)\Pr(Y=y)}=&\frac{1-F_X(x_{th})}{1-F_{X,exp,i}(x_{th})}\cdot\int_{x_{th}}^{+\infty}{f_{S|X}(s|x)f_{X,exp,i}(x)\mathrm{d}x}\\\nonumber&+\frac{F_X(x_{th})}{F_{X,exp,i}(x_{th})}\cdot\int_{-\infty}^{x_{th}}{f_{S|X}(s|x)f_{X,exp,i}(x)\mathrm{d}x}.
\end{align}
The posterior probabilities $\Pr(Y|s)$ and $\Pr(F|s)$ are similarly computed with Eqs.~\eqref{Eq:Posterior_condition} to~\eqref{Eq:Posterior_failure}.

The first experimental design ED1$=f_{X,exp,1}(x)$ corresponds to the distribution of the condition in this application $f_X(x)$ defined in Eq.~\eqref{Eq:True_dist}. Eq.~\eqref{Eq:Mod_Evidence} shows that in this case the model evidence $p_{exp,1}(s)$ and the posterior probabilities coincide with the model evidence and posterior probabilities obtained by applying Model (1) and the continuous/continuous likelihood $f_{S|X=x}(s)$. $f_{{S|Y},exp,1}(s)$ is therefore the true likelihood and provides the correct evaluation of expected costs.
This case is examined in Sections~\ref{SubS:Ex3_ROC} and~\ref{SubS:Ex3_PoDPFA}, and is called the reference case here.

When the experimental design differs from the distribution in the application, i.e., when $f_{X,exp}(x) \neq f_{X}(x)$, the model evidence and posterior probabilities deviate from the reference case and the conditional expected costs calculated with Eq.~\eqref{Eq:Exp_Cost_action} are incorrect.
This in turn affects the results of the decision problem and changes the optimal action attributed to each observation outcome by Eq.~\eqref{Eq:Sol_oneStep_cont_Cond}. It also impacts the expected cost calculated with Eq.~\eqref{Eq:Exp_cost}.

\begin{figure}[!h]
\centering
\includegraphics[width=0.7\linewidth]{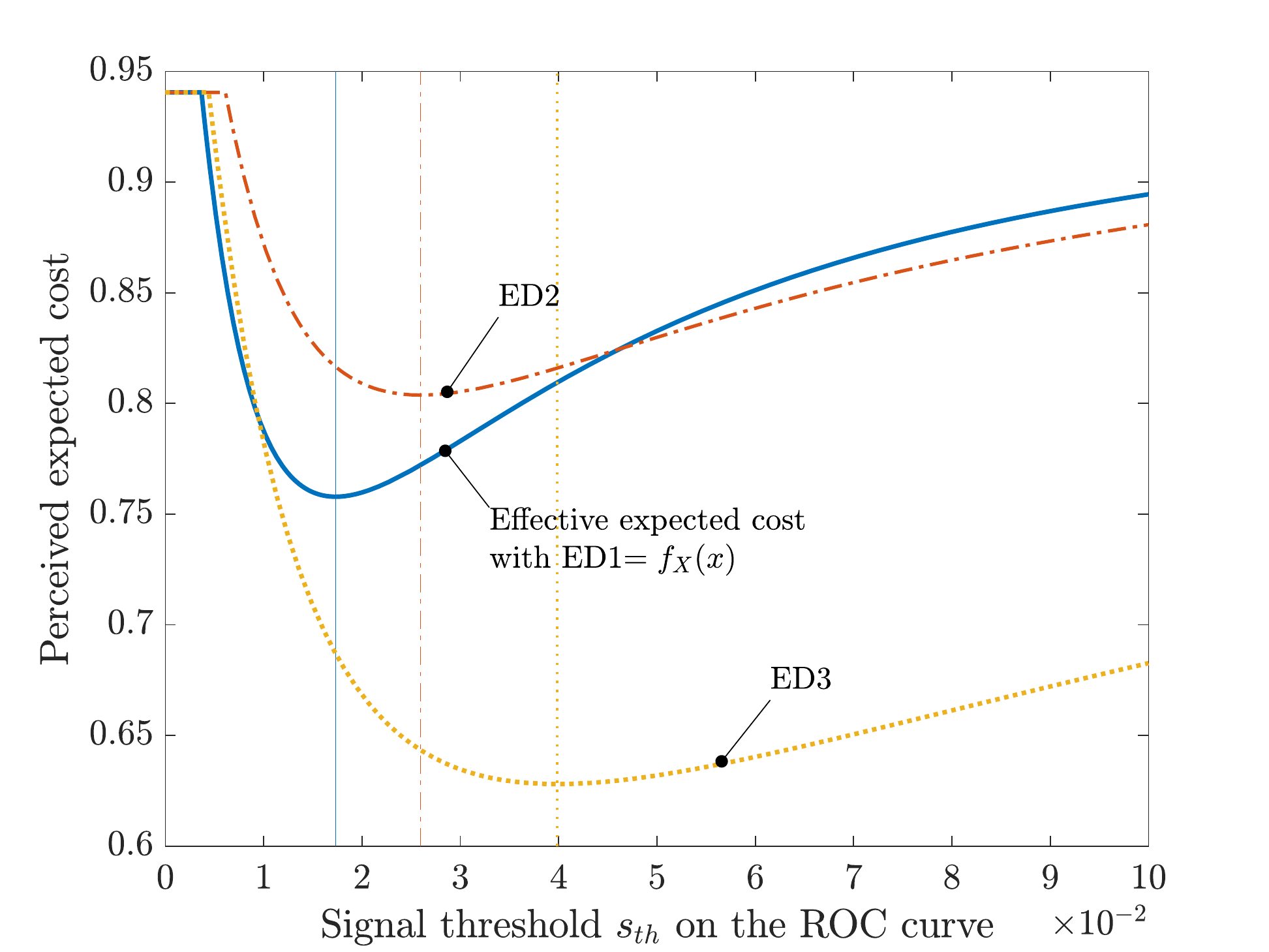}
\caption{Expected cost calculated for experimental designs ED1, ED2, and ED3. ED1 corresponds to the distribution of the condition in this application $f_X(x)$. ED2 and ED3 differ from $f_X(x)$. For a fixed threshold $s_{th}$, only ED1 provides the correct values for $PoD$ and $PFA$ and the associated expected cost. ED2 and ED3 give wrong $PoD$ and $PFA$, thus the perceived expected cost associated with $s_{th}$ is also wrong. The vertical lines of corresponding line style locate the perceived optimal threshold $s_{th,exp,i}$ for each experimental design, where optimal action $a_{opt}$ is $a_0$ if $s<s_{th,exp,i}$ and $a_R$ otherwise. The effective expected cost for these thresholds are read on the reference curve (ED1): The perceived expected cost for the optimal threshold for ED2 overestimates the effective expected cost; the perceived expected cost for the optimal threshold for ED3 underestimates it (see Table~\ref{Tab:Effect_exp_design}).}
\label{Fig:Effect_exp_design}
\end{figure}

One can evaluate this impact by overlaying the corresponding ROC curves on Fig.~\ref{Fig:Ex3_Exp_cost_ROC} and by plotting the expected cost and optimal actions of the decision problem with binary observation $\{I=1\}=\{S>s_{th}\}$ as a function of the fixed signal threshold $s_{th}$, as in Fig.~\ref{Fig:Effect_exp_design}.

Fig.~\ref{Fig:Effect_exp_design} shows that each experimental design gives a different evaluation of the expected cost for a given $s_{th}$. The constant part of each curve corresponds to the portion of the ROC curve in zone (a), and the other part falls in zone (b) (see Fig.~\ref{Fig:Ex3_Exp_cost_ROC}). These actions zones differ for each ROC curve.
For example, when the ROC curve is learned from experiments with ED2$=f_{X,exp,2}(x)$, one finds that imposing a threshold $3.5\cdot10^{-3}<s_{th}<6\cdot10^{-3}$ leads to the action "do nothing," whatever the outcome $s$. In contrast, under the reference case the optimal action is to repair if the signal is higher than the fixed threshold, and do nothing otherwise. By misevaluating the conditional expected costs, the experimental design leads to suboptimal actions.

Another problem is that the optimal calibration, in the form of a threshold $s_{th,exp,i}$, minimizes the wrong expected cost function. The effective expected cost at this threshold, obtained with the reference case, can in fact be higher or lower. In Fig.~\ref{Fig:Effect_exp_design}, the optimal threshold for ED2 is $s_{th,exp,2}=2.6\cdot10^{-2}$. At this point, the  effective expected cost given by ED1 is lower than the perceived expected cost for ED2.
Table~\ref{Tab:Effect_exp_design} summarizes these results for all investigated likelihoods.
\begin{table}[!h]
\caption{Effect of the experimental design on recommended actions and on the assessment of expected cost.}
\renewcommand{\arraystretch}{0.8}
\begin{tabular}{c |c |c |c}
Experimental design & Optimal threshold & Perceived expected cost & Effective expected cost\\
\hline
$f_{X,exp,1}(x)$ -- Reference & $1.7\cdot10^{-2}$& $0.76$& $0.76$\\
$f_{X,exp,2}(x)$ & $2.6\cdot10^{-2}$& $0.80$& $0.77$\\
$f_{X,exp,3}(x)$ & $4.0\cdot10^{-2}$& $0.63$& $0.81$\\
\end{tabular}
\label{Tab:Effect_exp_design}
\end{table}

In summary, ignoring the underlying experimental design of a given ROC curve can lead to suboptimal decisions and to an NDE system that is not used to its full potential. Additionally, the expected costs can be over- or underestimated, hence they affect the perceived VoI. Ultimately, this can lead the decision maker to favor one or another NDE system based on an erroneous evaluation of the NVoI of the NDE systems.

\section{Example 2: Half-cell potential measurement for corrosion detection}\label{SubS:Example_1and2}
\added{The purpose of this example is the investigation of the effects of different model choices on a real NDE technique. We consider the half-cell potential measurement of reinforcement corrosion in concrete structures, and we analyze the effect of choosing Model (3) or Model (4), with different calibration choices, on the outcomes of optimal decision making for a one-step and a two-step decision problem.}
\subsection{Description and model of the inspection technique}
Corrosion of reinforcement bars (rebars) in concrete structures is one of the leading deterioration mechanisms in civil infrastructure. Visible signs of corrosion on the concrete surface  occur when corrosion of the rebars is already extensive and major repairs are needed. This demonstrates the need for early detection of rebar corrosion.

As the rebars are encased in the concrete, NDE methods are required to monitor their condition. Half-cell potential measurement is such a method and it is used to detect whether corrosion has initiated.
This test measures the difference of electric potential between an electrode, directly connected to an exposed rebar, and a half-cell (reference electrode) placed on the concrete surface. The intensity of the chemical reaction responsible for corrosion at the half-cell is translated into a negative potential. The higher the amplitude of this electric potential, the higher the probability that corrosion has initiated in the rebar under the concrete cover \citep{Elsener_et_al_03}.

Empirical testing has revealed that the amplitude of the signal is clearly affected by environmental factors including humidity, chloride content, temperature, concrete cover, or concrete strength \citep[e.g.,][]{Lentz_et_al_02,Kessler_Gehlen_17}.
This is why obtaining relevant and accurate probabilistic models a priori is a challenge. In practice, the electric potentials are measured on a regular grid on the concrete structure and are summarized in a frequency plot. The measurement data is then classified into two sets, "corroded" and "not corroded," by applying a threshold on the potential. This is typically done by expert judgment. The downside of this method is that it relies on relatively few sample measurements 
and misclassification can occur at a high rate.

Some studies have proposed to model the performance of this NDE method with a non-site-specific likelihood \citep{Lentz_et_al_02,Faber_Sorensen_02}. Typically, the chosen model involves a continuous signal (the potential) and a binary measured condition (corrosion initiated/no corrosion), thus follows Model (3) described in Section~\ref{SubS:Mon_type}. However, these distributions are not used as such to perform Bayesian analysis. Instead, a threshold value on the potential is proposed, and the continuous signal transformed into a binary one \citep{Lentz_et_al_02}. Interestingly, the ROC representation is not common in this domain, even if it highlights best how the $PoD$ and $PFA$ vary with the chosen threshold.

\begin{figure}[!h]
\centering
\begin{subfigure}{0.5\linewidth}
\centering
\includegraphics[scale=0.5]{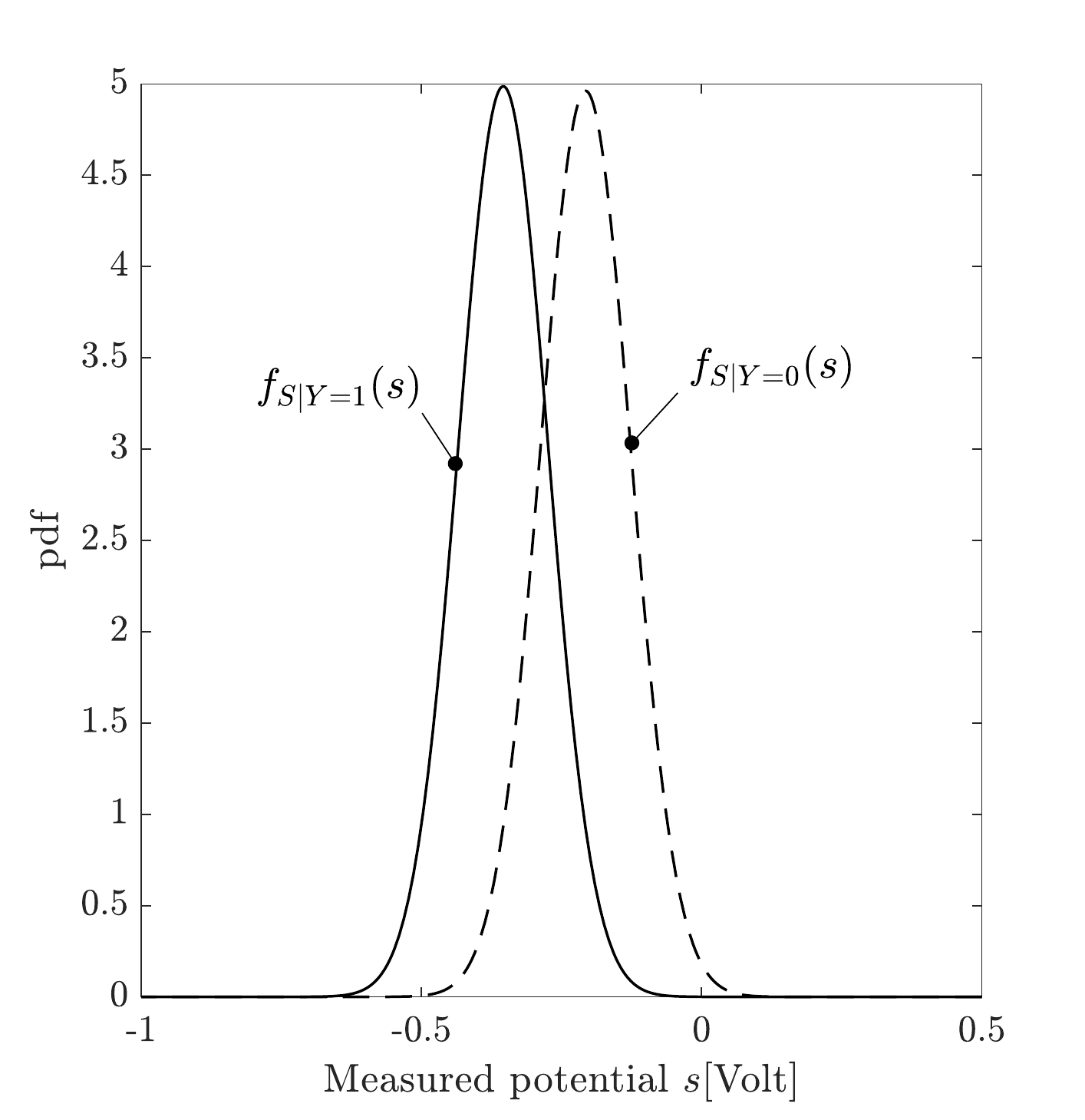}
\subcaption{}
\label{Fig:Ex_1_pdfs}
\end{subfigure}%
\begin{subfigure}{0.5\linewidth}
\centering
\includegraphics[scale=0.5,trim=30 0 30 0,clip]{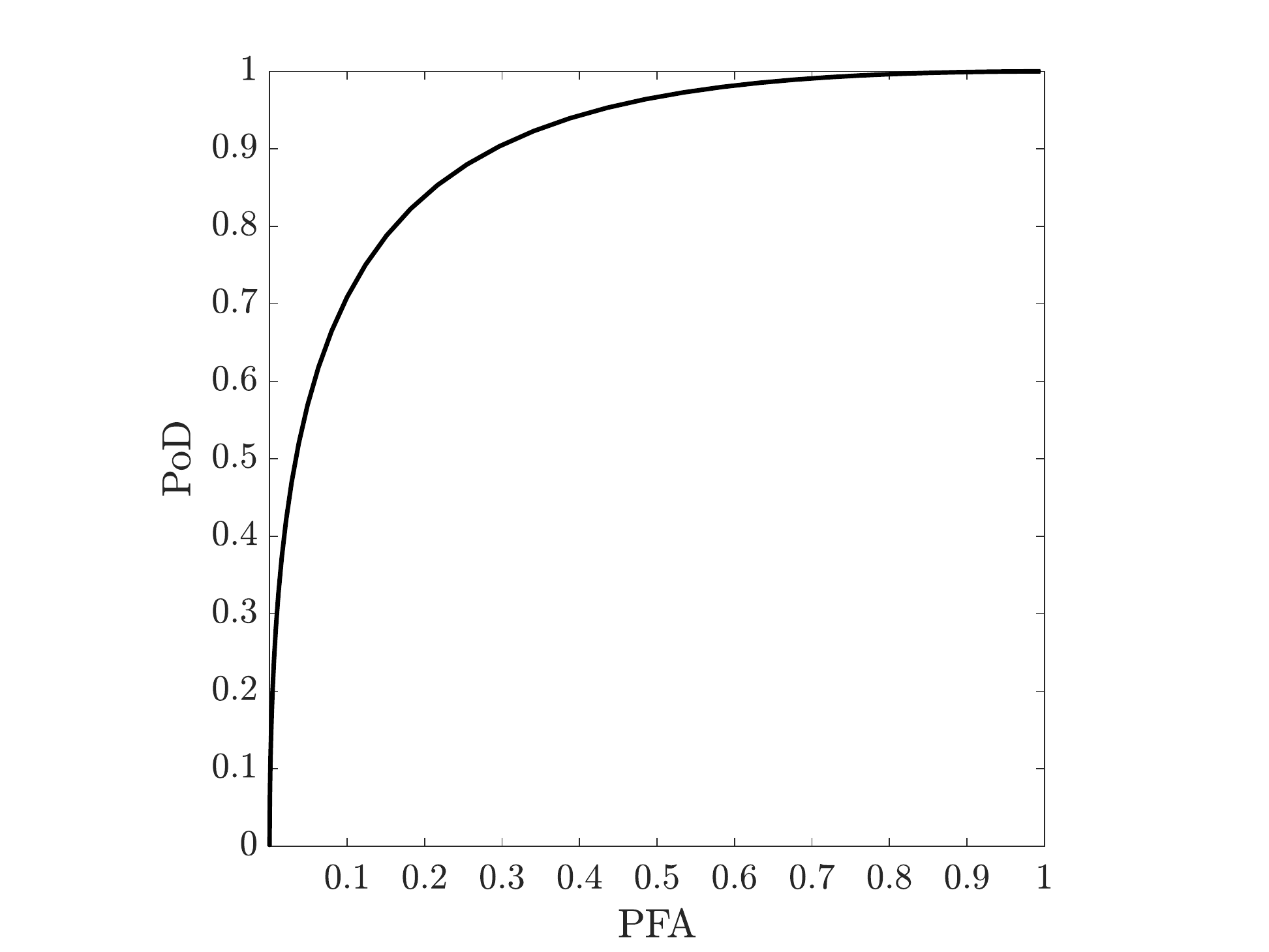}
\subcaption{}
\label{Fig:Ex_1_ROC}
\end{subfigure}
\caption{(a):The two likelihoods are depicted. (b): Corresponding ROC curve.}
\end{figure}

This example uses the NDE quality model proposed by \cite{Faber_Sorensen_02}. 
The absence and presence of corrosion is represented by $Y=0$ and $Y=1$, respectively.
We adopt normal pdfs for the likelihood after \citep{Faber_Sorensen_02}. 
The parameters of the normal distribution of the measured electric potential $S$ given that no corrosion has initiated ($Y=0$) are $\mu_0=-0.207$[Volt] and $\sigma_0=0.0804$[Volt]. The parameters of the normal distribution given that corrosion has initiated ($Y=1$) are $\mu_1=-0.354$[Volt] and $\sigma_1=0.08$[Volt]. The distributions are shown in Fig.~\ref{Fig:Ex_1_pdfs} and the corresponding ROC curve in Fig.~\ref{Fig:Ex_1_ROC}.
These distributions are (where $\varphi$ is the standard normal) pdf:

\begin{align}
\label{Eq:pdf_healthy}f_{S|Y=0}(s)=\mathcal{L}(Y=0;s)=\varphi\left(\frac{s-\mu_0}{\sigma_0}\right),\\\label{Eq:pdf_diseased}
f_{S|Y=1}(s)=\mathcal{L}(Y=1;s)=\varphi\left(\frac{s-\mu_1}{\sigma_1}\right).
\end{align}

We first investigate the one-step decision problem, where a repair action is taken based on the NDE result. In the second part, we consider a two-step decision problem, where information about the condition of the rebar is collected in two separate instances and repairs can be carried out each time. For both problems, we find the optimal course of action considering the continuous signal with Model (3). We show that fixing a threshold for Model (4), irrespective of the specific application is not optimal, for instance the one recommended by \cite{Faber_Sorensen_02}. We demonstrate how this fixed threshold can be optimized for both problems. 



\subsection{One-step decision problem}\label{SubS:Example_1}
\subsubsection{Decision model and a priori optimal action}
Corrosion of the system ($Y=1$) can eventually lead to failure. The decision maker has the opportunity to repair, $a_R$, at a cost $c_R=\text{\euro}5$ million, or to do nothing, $a_0$, at cost of \euro$0$. If the system is repaired, there are no further consequences. 
If the system is corroded ($Y=1$) but not repaired, the consequences are $c_{F}=\text{\euro}50$ million. If the system is not corroded ($Y=0$), there are no consequences.
The prior probability of corrosion is $\Pr(Y=1)=0.05$. 

%

The results are obtained following Section~\ref{App:Sol_one_step}. Without inspections, the expected total cost conditional on action $a_0$ is $c_{F}\Pr(Y=1)=\text{\euro}2.5$ million. Conditional on action $a_R$, it is $c_R=\text{\euro}5$ million.
Hence, the a priori optimal action is $a_{e_0}=a_0$, do nothing.


\subsubsection{Optimal a posteriori action considering the continuous signal}\label{SubS:Ex1_Cont_sign}
Eq.~\eqref{Eq:Sol_oneStep_Bin_Cond} applied with $Z=S$ is here adapted to
\begin{equation}\label{Eq:policy_ROC_2}
a_{opt}(s)=a_R \iff \frac{\mathcal{L}(Y=1;s)}{\mathcal{L}(Y=0;s)}>\frac{c_R\Pr(Y=0)}{(c_{F,0}-c_{F,R}-c_R)\Pr(Y=1)}=2.11.
\end{equation}
We derive
\begin{equation}
a_{opt}(s)=a_R \iff \left(\frac{s-\mu_0}{\sigma_0}\right)^2-\left(\frac{s-\mu_1}{\sigma_1}\right)^2>2 \log\left(\frac{\sigma_1}{\sigma_0}\frac{1}{2.11}\right).
\end{equation}
Finally,
\begin{equation}\label{Eq:Sol_ROC_Ex1}
a_{opt}(s)=a_R \iff s\in[s^*_1=-29.72\text{[Volt]};s^*_2=-0.31\text{[Volt]}].
\end{equation}

We see here that the signal domain is divided into three action zones. This is due to a mathematical artifact caused by the fact that the likelihoods are normal pdfs with unequal (although very similar) standard deviations. In practice, it does not make sense to not repair when the observed potential is very negative. The unsuitability of normal likelihoods for some NDE systems has been highlighted by \cite{Green_Swets_66}. The left bound $s^*_1$ in Eq.~\eqref{Eq:Sol_ROC_Ex1} is, however, very far from the expected values of the signal $S$, and does not significantly affect the results.


The expected cost associated with the a posteriori optimal action is evaluated by numerical integration from Eq.~\eqref{Eq:Exp_cost} and is \euro$1.4$ million.
\begin{comment}
\begin{equation}
\mathbf{E}[C_T]_{ROC}=c_R\left(\alpha\int_{-\infty}^{+\infty}\mathbbm{1}_{\frac{f_{S|Y=1}(s)}{f_{S|Y=0}(s)}<\gamma}\Pr(F|s,a_0)f_S(s)\mathrm{d}s+(1+\alpha  p_{F|R})\int_{-\infty}^{+\infty}\mathbbm{1}_{\frac{f_{S|Y=1}(s)}{f_{S|Y=0}(s)}>\gamma}f_S(s)\mathrm{d}s\right)
\end{equation}
\end{comment}


The VoI of this NDE system is $2.5-1.4=\text{\euro}1.1$ million, equivalent to a gain of $45\%$ relative to the a priori expected cost.
\subsubsection{Optimal a posteriori action for a binary signal with decision threshold $s_{th}$} 
Here, the continuous signal $S$ is processed to a binary outcome $\{I=0\}=\{S>s_{th}\}$ (no detection) and $\{I=1\}=\{S<s_{th}\}$ (detection), i.e., $Z=I$, and $\mathcal{L}(Y=1; I=1)=PoD(s_{th})$ and $\mathcal{L}(Y=0; I=1)=PFA(s_{th})$. 


\begin{comment}
\begin{align}
p_Y(I=1)=\frac{PFAp_Y}{\Pr(I=1)}\\
p_Y(I=0)=\frac{(1-PFA)p_Y}{1-\Pr(I=1)}
\end{align}
with $\Pr(I=1)=PFA(1-p_Y)+PoDp_Y$.

We get $\Pr(F|I,a_0)$ from Eq.~\eqref{Eq:Post_PoF}.

We can similarly establish the condition for the optimal choice of action, based on the outcome $I=1$ or $I=0$.
\end{comment}
As in Eq.~\eqref{Eq:policy_ROC_2}, the optimal actions are determined by a condition on $PoD$ and $PFA$:

\begin{align}
a_{opt}(I=1)=a_R \iff \frac{PoD}{PFA}>2.11, \\
a_{opt}(I=0)=a_R \iff \frac{1-PoD}{1-PFA}>2.11.
\end{align}


\begin{comment}
The optimal expected total cost for a fixed $PFA$ and $PoD$ is

\begin{align}
\mathbf{E}[C_T]_{\{PFA, PoD\}}=c_R\biggl(&\Pr(I=1)\left(\alpha\mathbbm{1}_{PFA>\gamma PoD}\Pr(F|I=1,a_0)+(1+\alpha p_{F|R})\mathbbm{1}_{PFA<\gamma PoD}\right)\nonumber\\&+\Pr(I=0)\left(\alpha\mathbbm{1}_{(1-PFA)>\gamma(1-PoD)}\Pr(F|I=0,a_0)+(1+\alpha p_{F|R})\mathbbm{1}_{(1-PFA)<\gamma(1-PoD)}\right)\biggr)
\end{align}
\end{comment}

\begin{figure}[!h]
\centering
\includegraphics[width=0.7\linewidth]{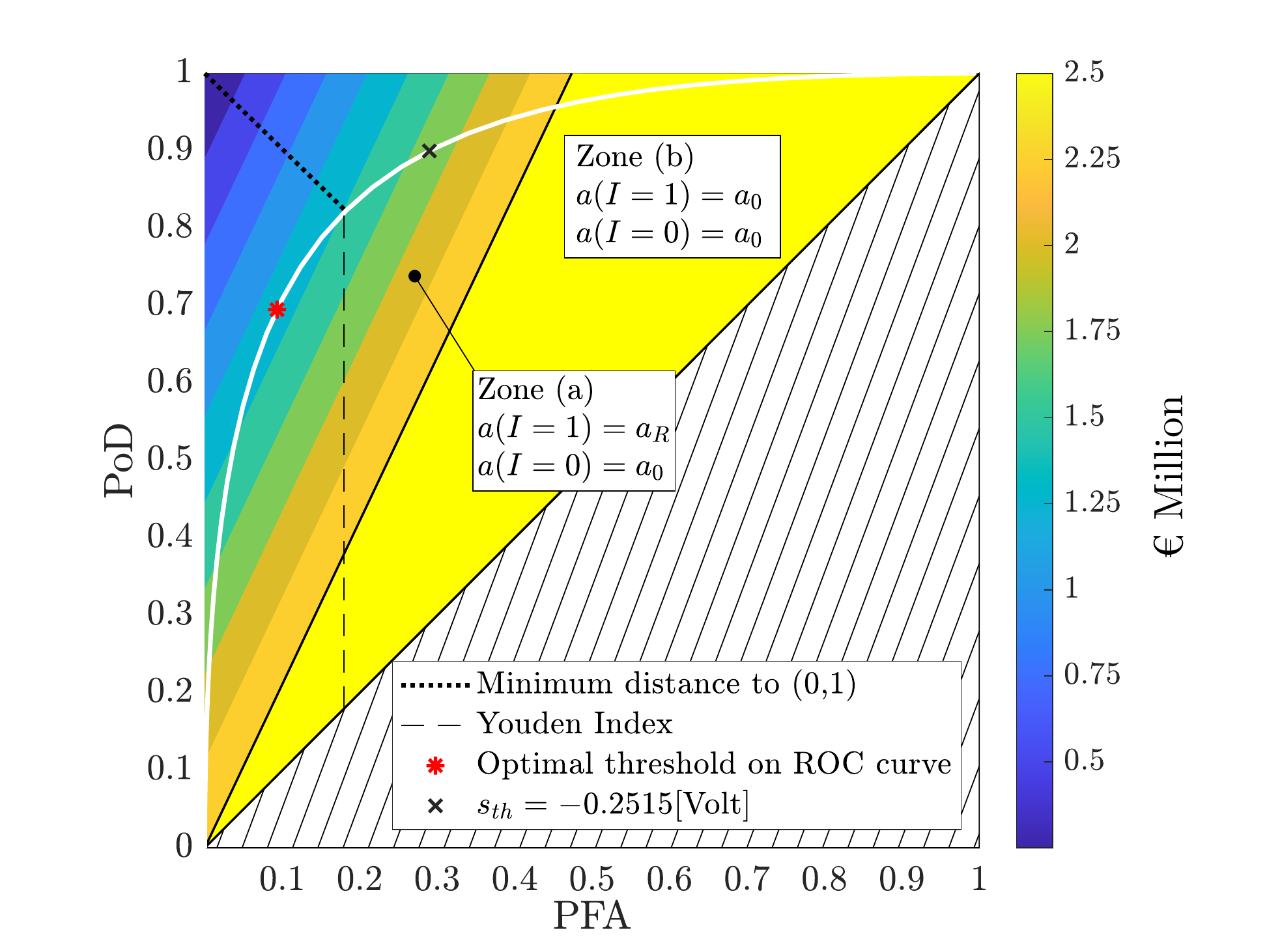}
\caption{Expected cost as a function of $PFA$ and $PoD$, corresponding optimal actions, and the ROC curve in white.  In zone (b), the selected action is $a_0$, hence the VoI is $0$. The hatched area corresponds to a monitoring system where the significance of $I=0$ and $I=1$ are inverted. The optimal point on the ROC curve obtained by Bayesian analysis differs from the Youden Index threshold and the point closest to (0,1). As a comparison, the threshold $-0.2515$[Volt] recommended by \cite{Faber_Sorensen_02} is indicated.}
\label{Fig:Optim_ROC}
\end{figure}

Fig.~\ref{Fig:Optim_ROC} shows the expected cost given by Eq.~\eqref{Eq:Exp_cost} as a function of $PFA$ and $PoD$. It also represents the ROC curve obtained from the pdfs of Eqs.~\eqref{Eq:pdf_healthy} and~\eqref{Eq:pdf_diseased}.


Fig.~\ref{Fig:Optim_ROC} also locates the cut-off point of the Youden Index, $s_{th}=-0.28$[Volt], which coincides for this ROC curve with the point closest to the origin. These are suboptimal operating points with an associated expected cost of \euro$1.5$ million, which is higher than the optimal cost found with Model (1).

The optimal threshold on the ROC curve that minimizes the expected cost is $s_{th,fix}=-0.31$[Volt], with associated cost \euro$1.4$ million. We note that this threshold is equal to bound $s^*_2$ found in Section~\ref{SubS:Ex1_Cont_sign}.  For this setup, Model (3) and Model (4) achieve the same performance without loss of information if the threshold is optimized for the problem at hand.

We can also evaluate the NDE system with cut-off point $-0.2515$[Volt] that \cite{Faber_Sorensen_02} directly apply on the measured potential for the repair decision criteria. It is shown on Fig.~\ref{Fig:Optim_ROC}. The expected total cost associated to that threshold is \euro$1.9$ million, which is $34\%$ more than the cost found with Model (3). It still yields a non-zero VoI. 



\subsection{Two-step decision problem}\label{SubS:Example_2}
\subsubsection{Decision model and a priori optimal actions}\label{SubS:TwoStep_prior}
We now consider a two-step decision problem. The repair actions are taken at each step based on the NDE data obtained up to that step. Additionally, the actions affect the state of the system at the next step.
The initial condition of the system is $Y_1$. The effect of the action $A_1\in\{a_0,a_R\}$ at time step $1$ on the state of the system is reflected in $Y_1'$. The condition of the system can evolve from $Y_1'$ to $Y_2$ at time step $2$. Both $Y_1$ and $Y_2$ are binary, with states "corroded" or "not corroded." Similarly, an action $A_2$ is taken that affects the state of the system, resulting in $Y_2^{'}$.
The immediate consequence of the concrete being corroded at time step $i$ is $C_{F,i}(Y_i'=1)=c_F$, where $c_F=\text{\euro}50$ million. If $Y_i'=0$, there are no associated consequences, i.e., $C_{F,i}(Y_i'=1)=\text{\euro}0$. The action cost is $C_{A,i}(a_R)=c_R=\text{\euro}5$ million and $C_{A,i}(a_0)=\text{\euro}0$.


The random variable $Y_1$ is characterized by $\Pr(Y_1=1)=0.1$. The conditional probabilities reflecting the decision taken at time step $i$ are given by $\Pr\left(Y_i'|A_i=a_0,Y_i\in\{0,1\}\right)=\mathbbm{1}_{Y_i'=Y_i}$, where $\mathbbm{1}$ is the indicator function, and $\Pr\left(Y_i'=1|A_i=a_R,Y_i\in\{0,1\}\right)=0$. The transition from $Y_1'$ to $Y_2$ is expressed by the conditional probability in Table~\ref{Tab:Ex_2_CPT}.

\begin{table}[!h]
\centering
\caption{Conditional probability $\Pr(Y_2|Y_1')$.}
\renewcommand{\arraystretch}{0.8}
\begin{tabular}{c |c| c}
&$Y_1'=0$ & $Y_1'=1$\\
\hline
$Y_2=0$ & $0.95$ & $0$\\
$Y_2=1$ & $0.05$& $1$\\
\end{tabular}
\label{Tab:Ex_2_CPT}
\end{table}

\begin{figure}[!h]
\begin{subfigure}{1\linewidth}
\centering
\includegraphics[scale=0.4]{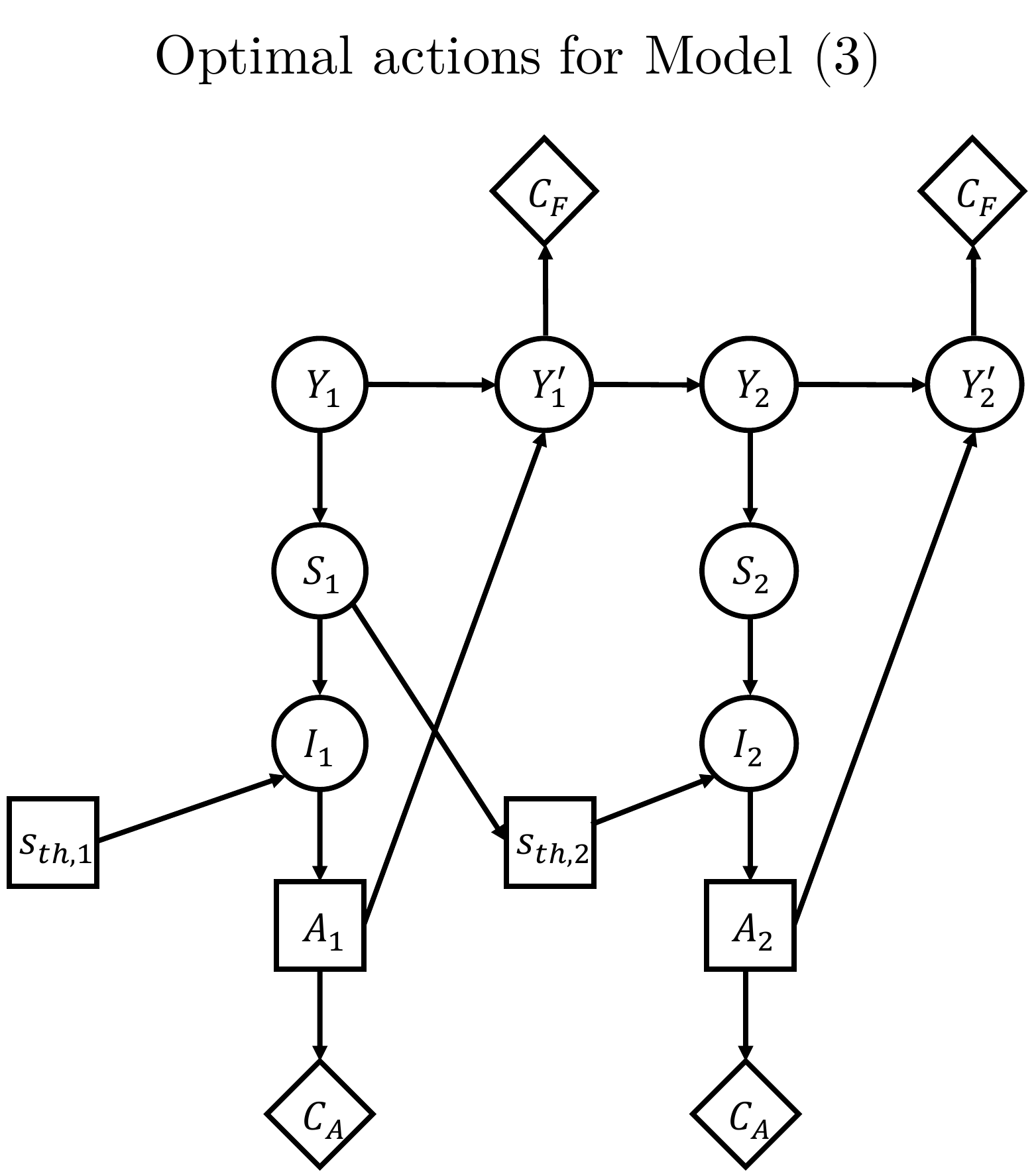}
\subcaption{}
\label{Fig:Two_thresholds}
\end{subfigure}
\begin{subfigure}{0.5\linewidth}
\centering
\includegraphics[scale=0.4]{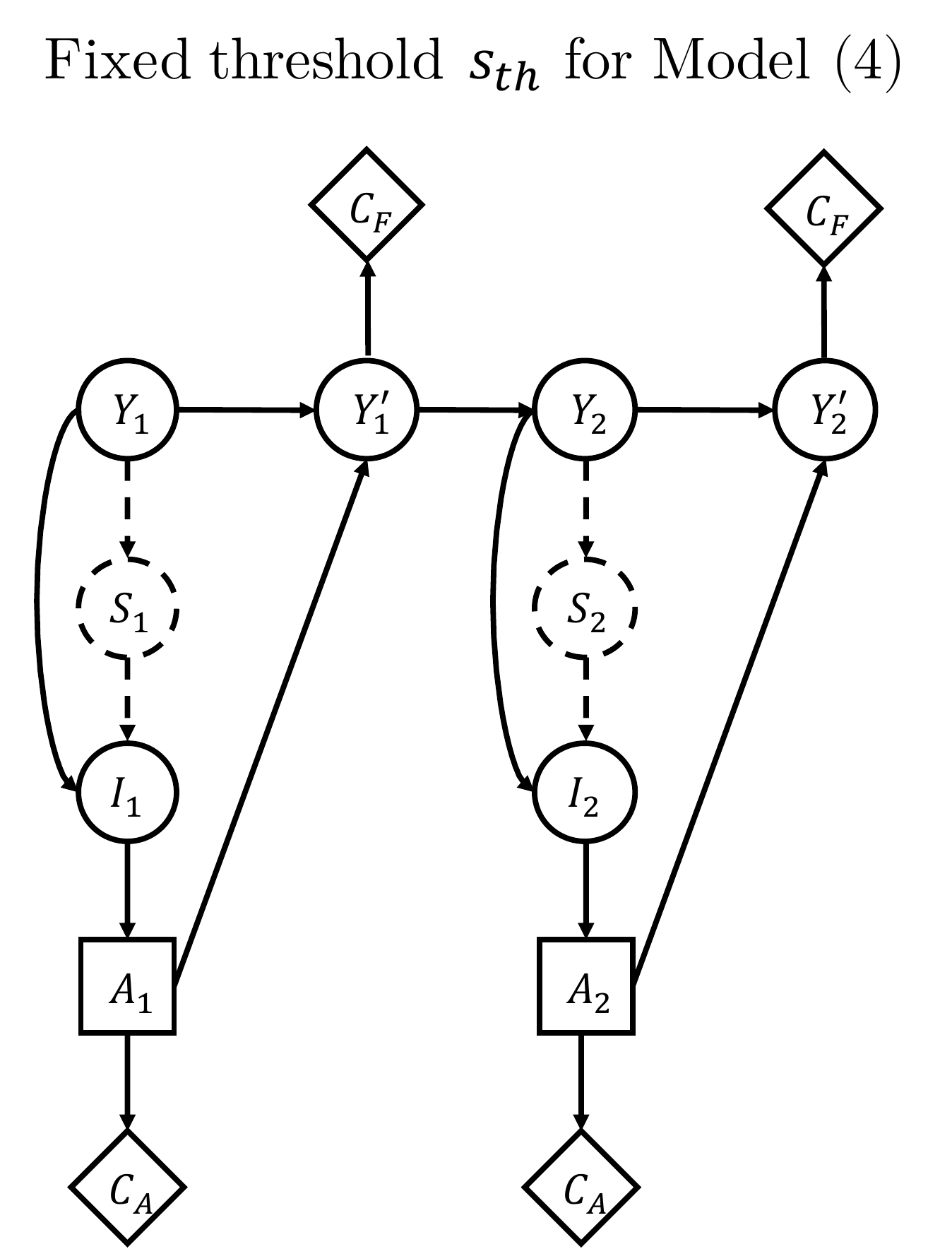}
\subcaption{}
\label{Fig:Fix_threshold_noOpt}
\end{subfigure}
\begin{subfigure}{0.5\linewidth}
\centering
\includegraphics[scale=0.4]{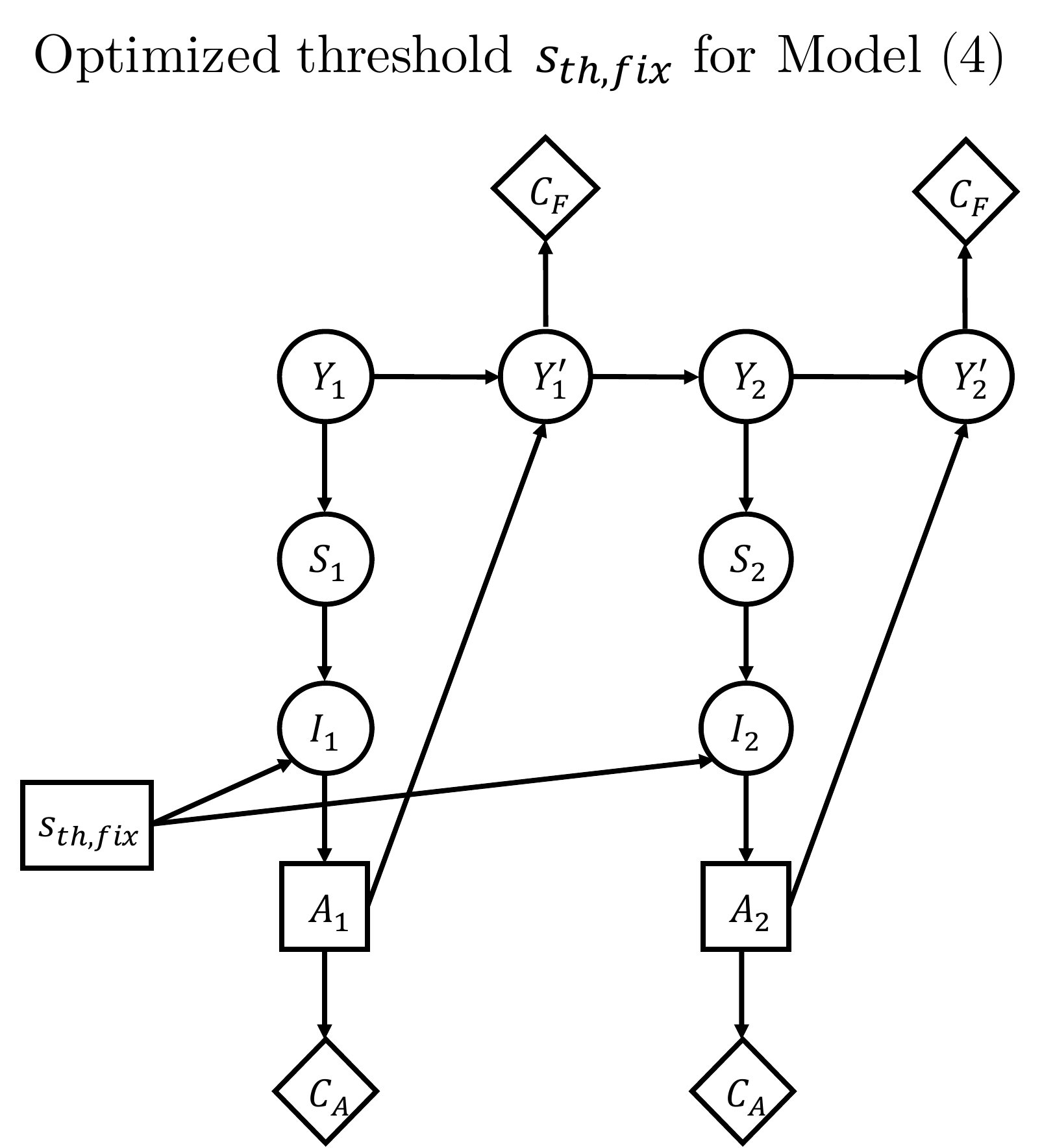}
\subcaption{}
\label{Fig:Fix_threshold}
\end{subfigure}
\caption{Influence diagrams of the two-step decision problem where the quality of NDE is described: (a) by Model (3), where the thresholds $s_{th,i}$ such that $\{I_i=1\}=\{S_i<s_{th,i}\}$ are optimized sequentially; (b) by Model (4) with a given threshold defining the likelihood $\Pr(I|Y)$; and (c) by Model (4), where the fixed threshold is optimized for the specific application.}

\end{figure}

We solve the two-step decision problem assuming three models of NDE quality. First, the NDE quality is described by Model (3) (Fig.~\ref{Fig:Two_thresholds}) and the thresholds and actions are optimized sequentially. The second quality model is Model (4), where a fixed point on the ROC curve in Fig.~\ref{Fig:Ex_1_ROC} is given (Fig.~\ref{Fig:Fix_threshold_noOpt}). Finally, we optimize the point on the ROC curve for this specific application (Fig.~\ref{Fig:Fix_threshold}). 
The details of the derivations are omitted here but are included as Supplementary Material. 

The optimal actions based only on prior knowledge are to first repair at time step $1$, and then do nothing at time step $2$, i.e., $(a_{1}, a_{2})_{e_0}=(a_R,a_0)$. The associated expected cost is \euro$7.5$ million. 

\subsubsection{Optimal a posteriori actions considering the continuous signal}\label{SubS:TwoStep_cont}
The actions are first optimized sequentially, based on the continuous half-cell potential measurements $S_1$ and $S_2$ at time step $1$ and $2$, respectively. This means that the NDE method is represented by Model (3).

This multi-step sequential decision problem with NDE results is in general difficult to resolve. A heuristic approach to the optimization can provide an approximate solution \citep{Bismut_Straub_19}. For this two-step problem, an analytical solution is obtained by backward induction \citep{Bellman_57a}. 

\begin{figure}[!h]
\centering
\includegraphics[width=0.7\linewidth]{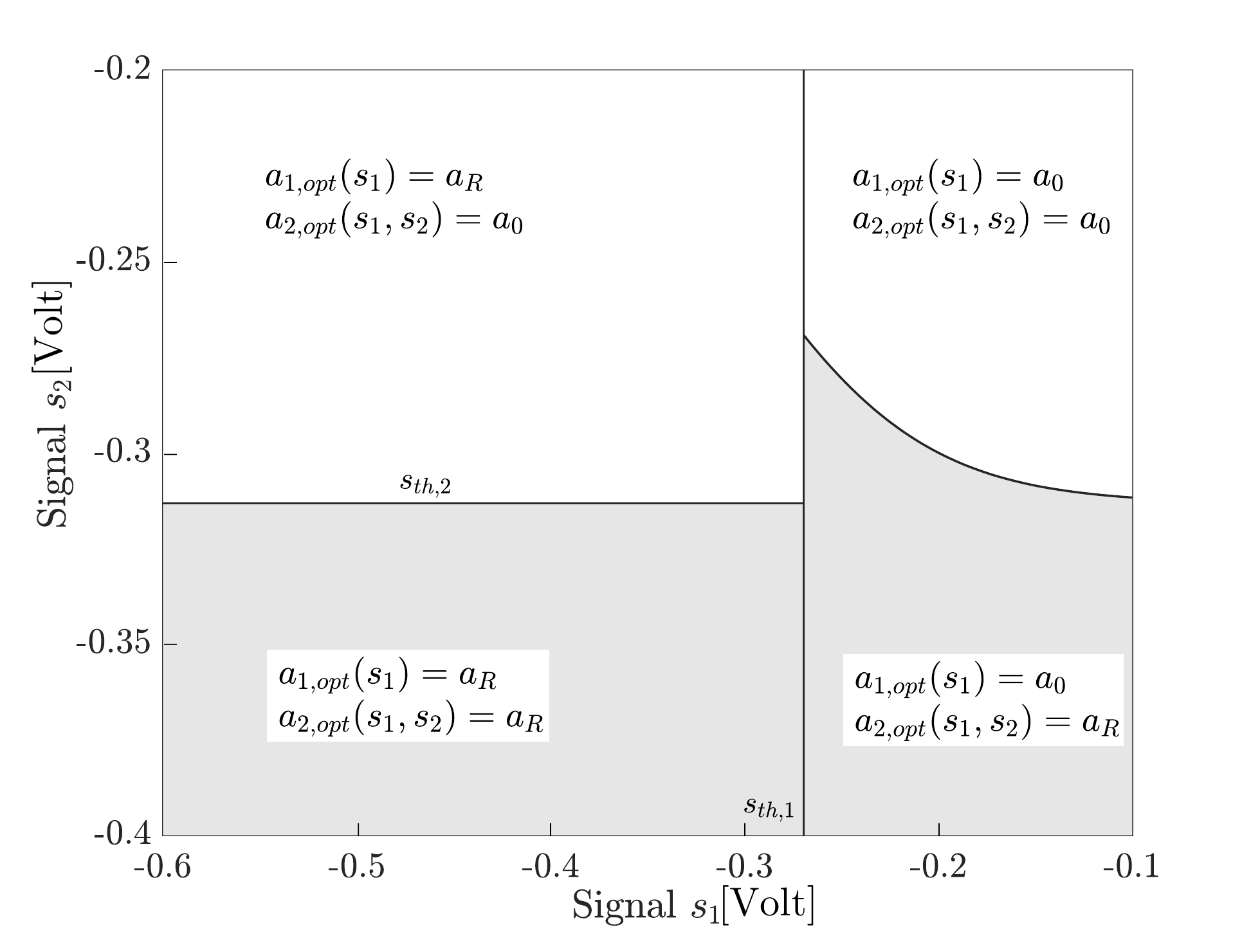}

\caption{Optimal policy given monitoring history $S_1$ and $S_2$. The optimal threshold on signal $S_1$ is $s_{th,1}=-0.27$[Volt]. The optimal threshold $s_{th,2}$ on signal $S_2$ depends on the observation $S_1$.}
\label{Fig:Opt_policy_Ex2}
\end{figure}

The optimal actions are shown in Fig.~\ref{Fig:Opt_policy_Ex2}, as a function of the measured signal $S_1$ and $S_2$ obtained sequentially. We can see that the optimal solution is equivalent to applying a threshold $s_{th,1}=-0.28$[Volt] on $S_1$, then a threshold $s_{th,2}$ on $S_2$ that depends on $S_1$, as illustrated by Fig.~\ref{Fig:Two_thresholds}. We see that if $S_1>s_{th,1}$, the optimal threshold at time step $2$ depends on the outcome at time step $1$. When the system is repaired at time step $1$, the optimal decision $a_{2,opt}$ is independent of outcome $S_1$.

The expected total cost of reacting optimally at each time step to the continuous signal is \euro$3.7$ million.
The VoI for this NDE system is $7.5-3.7=\text{\euro}3.8$ million. This result shows that considering the continuous signal is more optimal compared to classifying the signal according to a fixed rule.

\subsubsection{Optimal a posteriori actions for a binary signal with decision threshold}
In this configuration, the NDE system is described by Model (4). The corresponding influence diagram is drawn in Fig.~\ref{Fig:Fix_threshold_noOpt}. We investigate, for example, the fixed threshold $-0.2515$[Volt] from \cite{Faber_Sorensen_02}. The NDE system is described by the point $PoD=0.90$ and $PFA=0.29$ on the ROC curve (Fig.~\ref{Fig:Ex_1_ROC}). The expected total cost is \euro$4.2$ million and the VoI is $7.5-4.2=\text{\euro}3.3$ million. This fixed threshold $-0.2515$[Volt] is not optimal when compared to the VoI of Model (3) above.

\begin{figure}[!h]
\centering
\includegraphics[width=0.7\linewidth]{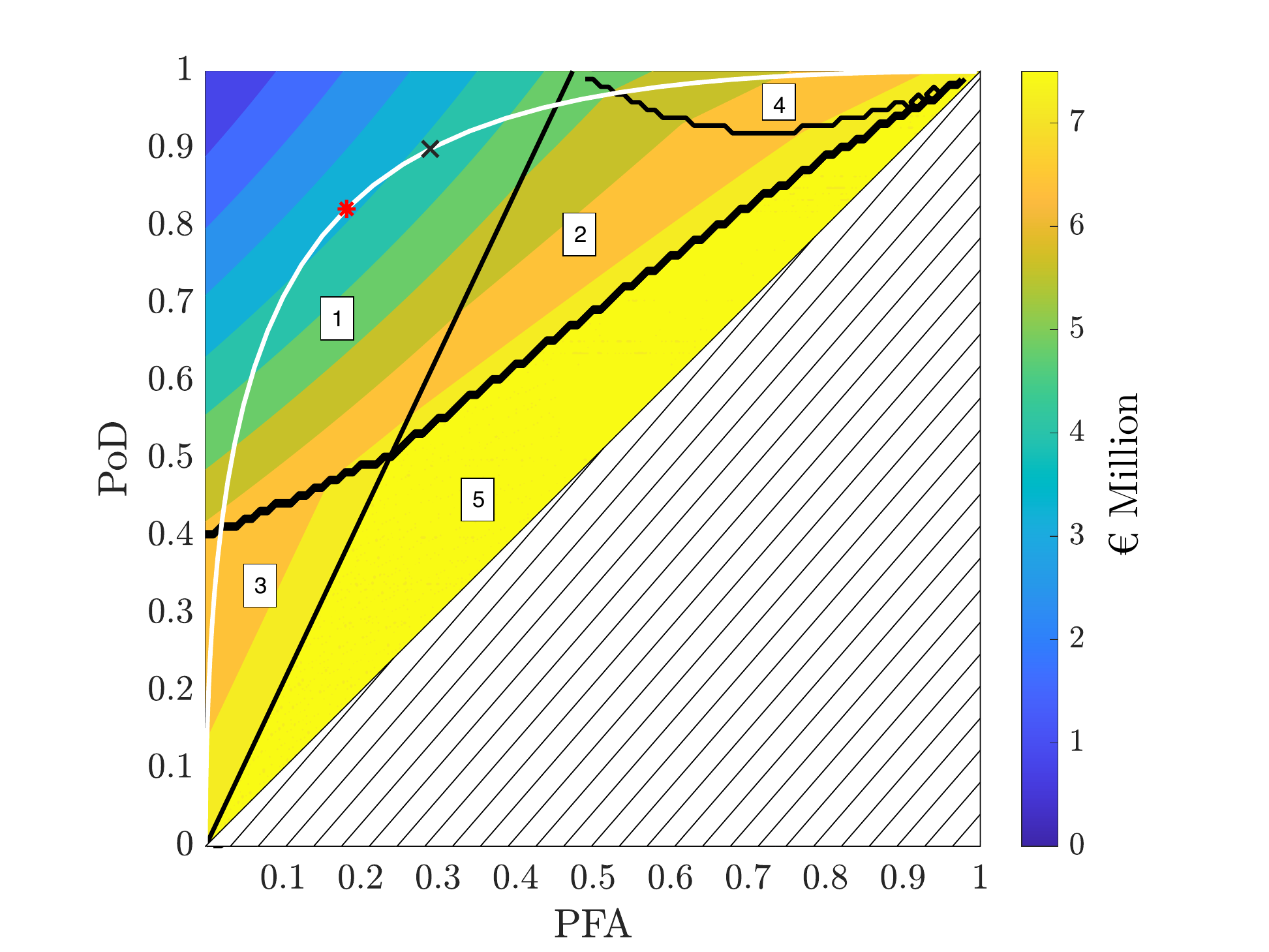}

\caption{ Expected total cost as a function of $PFA$ and $PoD$. The ROC curve is in white. The areas numbered 1 to 5 correspond to action zones. In zone 1, $a_{i,opt}(I_{i}=1)=a_R$, $a_{i,opt}(I_{i}=0)=a_0$; in zone 2, $a_{1,opt}(I_{1}=1)=a_R$, $a_{1,opt}(I_{1}=0)=a_0$, $a_{2,opt}(I_{2}=1,I_{1}=0)=a_R$, $a_{2,opt}(I_{2}=0,I_{1}=0)=a_0$, $a_{2,opt}(I_{1}=1)=a_0$; in zone 3, $a_{1,opt}=a_R$, $a_{2,opt}(I_{2}=0)=a_0$, $a_{2,opt}(I_{2}=1)=a_R$; in zone 4, $a_{1,opt}(I_{1}=1)=a_R$, $a_{1,opt}(I_{1}=0)=a_0$, $a_{2,opt}=a_0$; in zone 5, $a_{1,opt}=a_R$, $a_{2,opt}=a_0$. Zone 5 corresponds to $VoI=0$. The red asterisk locates the optimal fixed operating point with threshold $s_{th,fix}=-0.28$[Volt]. As a comparison, the cross locates the point for threshold $-0.2515$[Volt] from \citep{Faber_Sorensen_02}.}
\label{Fig:Cost_Fix_Threshold}
\end{figure}

Using Model (3) for the considered NDE method provides the optimal actions for the two-step decision problem. However, one can optimize the fixed threshold and point on the ROC curve for Model (4), as illustrated by the modified influence diagram of Fig.~\ref{Fig:Fix_threshold}.
To do so, we first determine sequentially the optimal actions $a_{1,opt}(I_1)$ and $a_{2,opt}(I_1,I_2)$, for any $PFA$ and $PoD$ values. The resulting expected cost is depicted in Fig.~\ref{Fig:Cost_Fix_Threshold}. The optimal threshold is found on the ROC curve that maximizes the expected cost along the curve. One finds $s_{th,fix}=-0.28$[Volt] and the associated expected cost is \euro$3.8$ million. The VoI for this NDE system is $7.5-3.8=\text{\euro}3.7$ million. The ROC curve and optimal operating point are also shown in Fig.~\ref{Fig:Cost_Fix_Threshold}. As a comparison, the previously evaluated threshold $-0.2515$[Volt] is also plotted on the ROC curve.



For given $PFA$ and $PoD$ values, the optimal actions of the two-step problem with Model (4) are defined by the zone, numbered 1 to 5, to which the $PFA$ and $PoD$ values belong in Fig.~\ref{Fig:Cost_Fix_Threshold}. In zone 1, the optimal actions are condition-based and memoryless, which corresponds to $a_{opt,i}(I_i=1)=a_R$ and $a_{opt,i}(I_i=0)=a_0$. This is not true when the $PFA$ and $PoD$ values belong to zones 2 to 5. It is therefore necessary to solve the decision problem, as any point on the ROC curve does not result in optimal actions that are condition-based and memoryless. This conclusion echoes the results by \citet{Bertovic_16}, who warns that the outcomes of mechanized NDE may be blindly trusted and interpreted as a call to action, to the detriment of a formal information assessment.

\begin{figure}[!h]
\centering
\includegraphics[width=0.7\linewidth]{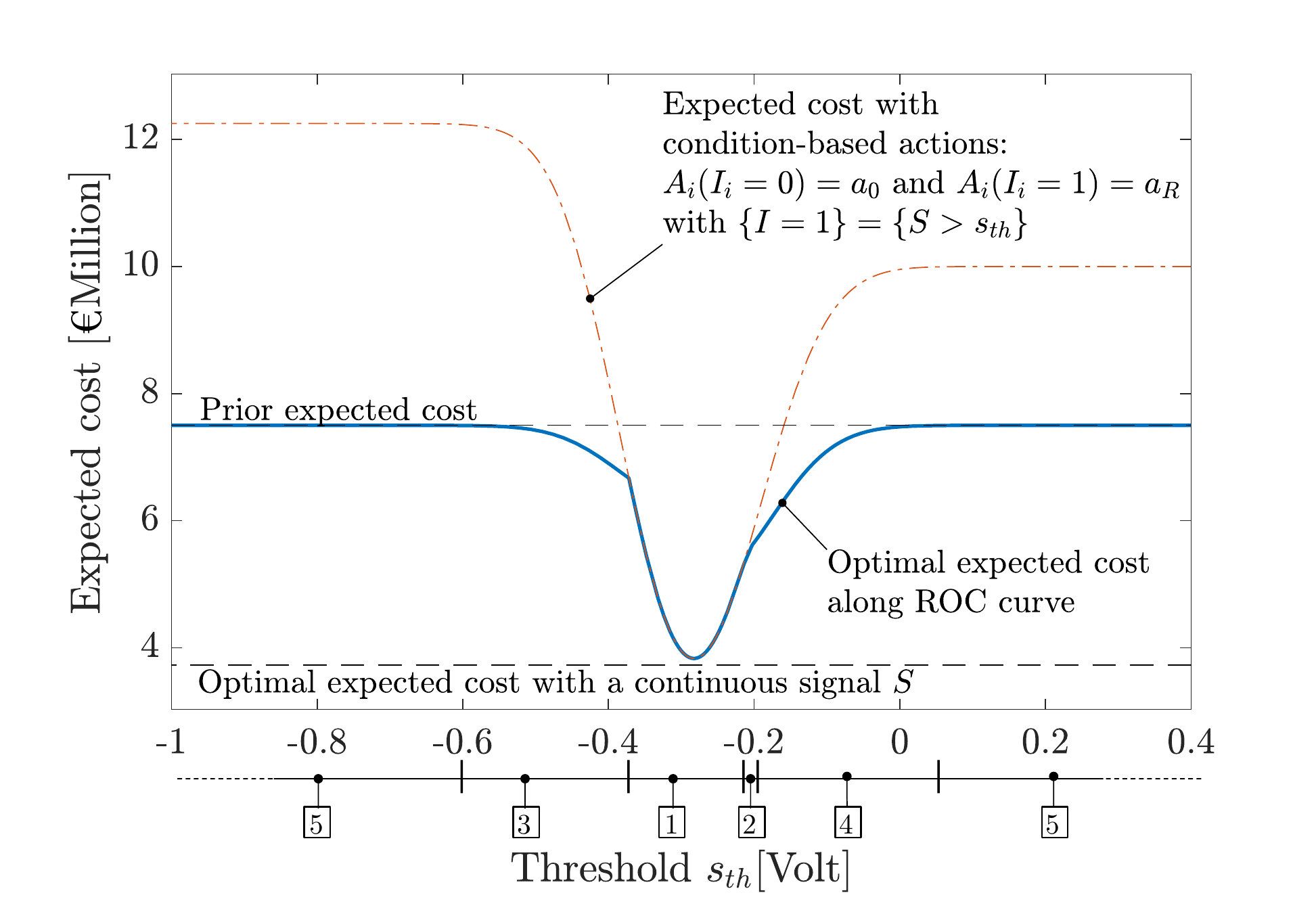}
\caption{Expected cost associated with a condition-based and memoryless maintenance strategy and with the optimal actions for Model (4). The action zones, 1 to 5, of Fig.~\ref{Fig:Cost_Fix_Threshold} are indicated as a function of the threshold value. For certain threshold values, the expected cost associated with the condition-based memoryless maintenance strategy is higher than the prior expected cost. }
\label{Fig:Comparison_condition_predictive}
\end{figure}

The effect of ignoring the different action zones is illustrated in Fig.~\ref{Fig:Comparison_condition_predictive}. This figure compares the optimal expected total costs along the ROC curve and the expected total cost of following a condition-based and memoryless maintenance strategy (as in zone 1). The curve corresponding to the optimal expected cost along the ROC curve stays within the lower and upper bounds, which are respectively the optimal expected cost with Model (3) at \euro$3.7$ million (Section~\ref{SubS:TwoStep_cont}) and the optimal a priori expected cost of \euro$7.5$ million (Section~\ref{SubS:TwoStep_prior}). In contrast, implementing a condition-based and memoryless maintenance strategy with a fixed NDE threshold, for example $-0.4$[Volt] in zone 3, results in an expected cost of \euro$8$ million. This cost is higher than the expected cost of \euro$6.9$ million associated with the optimal actions for Model (4) with this same threshold, and even higher than the a priori expected cost.

\section{Concluding remarks}
This paper reviews existing models NDE quality in the context of a unifying framework .

We have clarified the connection between the different models of NDE quality, such as PoD curves or ROC curves, through a base model in which both the condition and the NDE outcome are modeled continuously. 
The models with a binary observed signal or binary condition are derived by imposing thresholds on the observed signal and the condition. For models that use a  binary condition, one must also impose a distribution on the underlying continuous condition. 

In practice, there currently is no systematic approach to choose and calibrate NDE models and handle NDE data for reliability assessment or maintenance planning \citep{Kurz_et_al_12}. The presented framework can support the development and appropriate application of NDE models to real-life settings.
Ideally, the base model is learned from experimental data and any other model can then be derived from this base model. 
When the base model is not available, it can still be considered at an abstract level to appraise the quality of given NDE models. As we demonstrate, the use of given PoD curves or ROC curves are likely to yield suboptimal actions for a specific decision context, since the signal and condition thresholds are fixed without considering the decision problem.
The framework also highlights the importance of explicitly accounting for the experimental design conditions that underlie ROC curve models, which should be assessed within the decision context. 

Calibration of an NDE system in function of the decision settings (e.g., cost model) is beneficial. Such a calibration is often done implicitly. In particular, on-site inspectors tend to report a larger damage when the occurrence of failure is cost-critical, and vice-versa will be more cautious to diagnose damage when the costs of repair are very high \citep{daSilva_dePadua_12,Bertovic_16}. This paper derives the formal analysis associated with these decisions.
Bayesian decision analysis provides the means to compute the VoI, which allows direct comparison between NDE systems.
This analysis gives the opportunity to calibrate an NDE system to suit the decision parameters such as the cost of mitigating actions and the expected consequences of failure.
In Sections~\ref{SubS:Example_3and4} and~\ref{SubS:Example_1and2} we have evaluated the VoI in simple one-step and two-step decision problems. We find that using the model with continuous variables leads to the best decisions. In contrast, systematically associating NDE system outcomes with maintenance actions can lead to sub-optimal decisions and a detrimental use of the inspection data. 

Finding the VoI of NDE systems in sequential decision problems remains, however, a challenging  task. 
The formal and complete decision analysis for an NDE system applied to continuous monitoring is not tractable \citep{Papadimitriou_Tsitsiklis_87}. \replaced{As an alternative}{In this case}, a decision rule, or heuristic, can be applied for the treatment of information, and some heuristic parameters optimized such that the information from NDE system is collected, interpreted and exploited in a good manner \citep{Sheils_et_al_10,Bismut_Straub_19}.


\section*{Acknowledgments}
This research is supported by the Deutsche Forschungsgemeinschaft (DFG) through the TUM International Graduate School of Science and Engineering (IGSSE).

\section*{Supplementary Material}\label{Appendix}
The following supplementary material is available online at \url{https://arxiv.org/src/2103.12853v1/anc/Supplementary_Material.pdf}: \textbf{A}. Solution of the two-step decision problem of Example 2.


\bibliographystyle{elsarticle-harv}

\bibliography{Journal_Article_2_Feb2022.bbl}

\end{document}